%% file: T.tex
\documentclass{lmcs}
\usepackage{pstricks,pst-node}
\usepackage{enumerate}
\usepackage{url}

\newcommand {\defin}[1]{\emph{#1}}

\newenvironment{deflist}[1]%
{\begin{list}{}%
{\settowidth{\labelwidth}{#1}%
\setlength{\leftmargin}{\labelwidth}%
\addtolength{\leftmargin}{\labelsep}%
}}%
{\end{list}}

\newenvironment{deflisteng}[1]%
{\begin{list}{}%
{\settowidth{\labelwidth}{#1}%
\setlength{\leftmargin}{\labelwidth}%
\addtolength{\leftmargin}{\labelsep}%
\setlength{\topsep}{0pt}%
\setlength{\itemsep}{0pt}%
}}%
{\end{list}}%

\newcommand {\DL}{\begin{deflist}}
\newcommand {\EDL}{\end{deflist}}
\newcommand {\DLE}{\begin{deflisteng}}
\newcommand {\EDLE}{\end{deflisteng}}

\newcommand {\db}{\displaybreak[0]}

	\newcommand{\ie}{i.e.\ }
	\newcommand{\eg}{e.g.\ }
	\newcommand{\wrt}{w.r.t.\ }
	
	\newcommand{\resp}{resp.\ }
	\newcommand{\esp}{esp.\ }
	
        \newcommand{\famm}{fully abstract model}
        
 	\newcommand{\med}{\medskipamount}

\newcommand{\Berry}{\cite{Berry}}

\newcommand {\typ}{\colon} 
\newcommand {\fun}{\mathbin{\to}}    
\newcommand {\pro}{\times}    
\newcommand {\lpro}{\langle}    
\newcommand {\rpro}{\rangle}    
\newcommand {\proi}{\pi_1}    
\newcommand {\prot}{\pi_2}    

\newcommand {\la}{\lambda}

\newlength{\botw}
\newcommand {\eval}{\operatorname{eval}}

\newcommand {\lex}{\sqsubseteq} 
\newcommand {\Lubex}{\bigsqcup\nolimits} 

\newcommand {\down}{\mathord{\downarrow}}

\newcommand {\id}{\mbox{\textnormal{id}}}
\newcommand {\comp}{\circ}

\newcommand {\set}[2]{\{\, #1 \mid #2 \,\}}  
\newcommand {\ifthen}{\Longrightarrow}
\newcommand {\thenif}{\Longleftarrow}
\newcommand {\ifff}{\Longleftrightarrow}

\newcommand {\sub}{\subseteq}
\renewcommand {\Join}{\bigcup}
\newcommand {\all}[1]{\forall #1 . \;}
\newcommand {\exi}[1]{\exists #1 . \;}

\input{sazmacro}

\begin{document}
\title[From Sazonov's Natural Domains to Closed Directed-Lub Partial Orders]
{From Sazonov's Non-Dcpo Natural Domains to Closed Directed-Lub Partial Orders}
\author{Fritz M\"uller}
\address{Saarland University, Department of Computer Science, Campus E1.3, 66123 Saarbr\"ucken, Germany,
 \url{http://www.rw.cdl.uni-saarland.de/~mueller}}
\email{\texttt{($\la$x.muellerxcdl.uni-saarland.de)@} }

\input{T1}

\input{T2}

\input{T3}
\input{T4}

\input{T5}
\input{T6}
\input{T7}
\input{T8}
\input{T9}

\section*{Acknowledgement}
I thank Reinhold Heckmann for very carefully reading the drafts of this paper and many fruitful discussions
and hints. Proposition \ref{p:algfin} is due to him.\\
I thank Reinhard Wilhelm, Sebastian Hack and the members of their chair for their support,
\esp Roland Lei\ss a and Klaas Boesche for help with the computer.
\bibliography{bib}
\bibliographystyle{plain}
\end{document}

%% file: sazmacro.tex
\newcommand{\Saz}{\cite{Sazonov:natural}}
\newcommand{\preclosure}{preclosure}
\newcommand{\cont}{continuous}

\newcommand{\vphi}{\varphi}

\newcommand{\curry}{\operatorname{curry}}
\newcommand{\pre}{\operatorname{\mathit{pre}}}
\newcommand{\lab}{\operatorname{\mathit{lab}}}
\newcommand{\cl}{\operatorname{cl}}
\newcommand{\clD}{\operatorname{cl}_D}

\newcommand{\clE}{\operatorname{cl}_E}
\newcommand{\clP}{\operatorname{cl}_P}

\newcommand{\lubcomp}{\operatorname{lub-comp}}
\newcommand{\Rcomp}{\operatorname{\mathcal{R}-comp}}
\newcommand{\Rinc}{\operatorname{\mathcal{R}-inc}}
\newcommand{\Reta}{\operatorname{\mathcal{R}-\eta}}
\newcommand{\inn}{\operatorname{in}}
\newcommand{\innD}{\operatorname{in}_D}

\newcommand{\length}{\operatorname{length}}

\newcommand{\sazlub}{\biguplus}
\newcommand{\lexD}{\lex_D}
\newcommand{\lexE}{\lex_E}
\newcommand{\nlub}{\to}
\newcommand{\nlubD}{\nlub_D}
\newcommand{\nlubE}{\nlub_E}
\newcommand{\nlubP}{\nlub^P}
\newcommand{\lset}{\sqsubseteq}
\newcommand{\ruleto}{\mathord{\leadsto}}

\newcommand{\funad}{\rightharpoonup}
\newcommand{\downo}{\mathord{\down^0}}

\newcommand{\fDE}{f\typ D\fun E}
\newcommand{\Cel}{|C|}
\newcommand{\Del}{|D|}
\newcommand{\Eel}{|E|}
\newcommand{\Dcl}{\hat{D}}
\newcommand{\Dhat}{\hat{D}}
\newcommand{\Do}{D^0}
\newcommand{\DproEel}{|D\pro E|}
\newcommand{\elem}[1]{|#1 |}
\newcommand{\Xb}{\overline{X}}
\newcommand{\Bb}{\overline{B}}
\newcommand{\Cb}{\overline{C}}
\newcommand{\Cs}{C^{\ast}}
\newcommand{\Ib}{\bar{I}}
\newcommand{\Is}{I^{\ast}}
\newcommand{\cb}{\overline{c}}
\newcommand{\cnb}{\overline{c_n}}
\newcommand{\dnb}{\overline{d_n}}
\newcommand{\xiI}{\{x_i\}_{i\in I}}
\newcommand{\zjI}{\{z_j\}_{j\in I}}
\newcommand{\yij}{y_{ij}}
\newcommand{\yijIJ}{\{\yij\}_{i\in I\!, j\in J}}
\newcommand{\yijJ}{\{\yij\}_{j\in J}}
\newcommand{\yijII}{\{\yij\}_{i,j\in I}}
\newcommand{\yijIIj}{\{\yij\}_{j\in I}}
\newcommand{\yijIIi}{\{\yij\}_{i\in I}}
\newcommand{\yiiI}{\{y_{ii}\}_{i\in I}}
\newcommand{\aiset}{\{a_i\}_{i\geq 1}}
\newcommand{\biset}{\{b_i\}_{i\geq 1}}
\newcommand{\bipset}{\{b_i'\}_{i\geq 1}}
\newcommand{\ciset}{\{c_i\}_{i\geq 1}}
\newcommand{\diset}{\{d_i\}_{i\geq 1}}
\newcommand{\bij}{b_{ij}}
\newcommand{\biw}{b_{iw}}
\newcommand{\biwj}{b_{iwj}}
\newcommand{\cij}{c_{ij}}
\newcommand{\cik}{c_{ik}}
\newcommand{\ciw}{c_{iw}}
\newcommand{\cnw}{c_{nw}}
\newcommand{\ciwj}{c_{iwj}}
\newcommand{\ciwk}{c_{iwk}}
\newcommand{\Rcal}{\mathcal{R}}
\newcommand{\Scal}{\mathcal{S}}
\newcommand{\Ccal}{\mathcal{C}}

\newlength{\underlength}
\newlength{\sunderlength}
\newcommand{\screal}{\mathord{\mbox{{\makebox[0pt][l]{\makebox[\sunderlength]{$\scriptstyle\uparrow$}}}%
\raisebox{-3pt}{\mbox{$\scriptstyle -$}}}}}

\newlength{\arrowlength}
\newcommand{\under}{\mathrel{\mbox{{\makebox[0pt][l]{\makebox[\arrowlength]{$\sqsubset$}}}%
\raisebox{-5pt}{\mbox{$\leftarrow$}}}}}

\newcommand{\real}[1]{#1^{\screal}}
\newcommand{\breal}[1]{#1^{\uparrow}}
\newcommand{\Dreal}{\real{D}}
\newcommand{\Ereal}{\real{E}}

\newcommand{\Pot}{\mathcal{P}}
\newcommand{\Potlub}{\bar{\Pot}}
\newcommand{\Potdlub}{\Pot^\uparrow}
\newcommand{\join}{\cup}

\newcommand{\sups}{\supseteq}
\newcommand{\exiun}[1]{\exists ! #1 .\;}
\newcommand{\Right}{$\Rightarrow$}
\newcommand{\Left}{$\Leftarrow$}
\newcommand{\fset}[1]{#1^{+}}
\newcommand{\fsetf}{\fset{f}}

\newcommand{\fsset}[1]{#1^{++}}
\newcommand{\fssetf}{\fsset{f}}

\newcommand{\Kcat}{\mathbf{K}}

\newcommand{\Ccat}{\mathbf{C}}
\newcommand{\Rcat}{\mbox{$\mathcal{R}$-$\mathbf{cat}$}}
\newcommand{\Dcpo}{\mathbf{Dcpo}}
\newcommand{\Dlubpo}{\mathbf{Dlubpo}}
\newcommand{\Cdlubpo}{\mathbf{Cdlubpo}}
\newcommand{\Stdlubpo}{\mathbf{S10dlubpo}}
\newcommand{\Rdcpo}{\mathbf{Rdcpo}}
\newcommand{\Crdcpo}{\mathbf{Crdcpo}}
\newcommand{\terminal}{\top}
\newcommand{\expo}{\mathbin{\Rightarrow}}
\newcommand{\DexpoE}{D\expo E}
\newcommand{\DexpoEel}{|D\expo E|}
\newcommand{\DexpoKSE}{D\expo_{\Kcat}^{\ast}E}
\newcommand{\DexpoKSEel}{|D\expo_{\Kcat}^{\ast}E|}
\newcommand{\DexpoKE}{D\expo_{\Kcat}E}
\newcommand{\DexpoKEel}{|D\expo_{\Kcat}E|}
\newcommand{\DexpoSE}{D\expo^{\ast}E}
\newcommand{\DexpoSEel}{|D\expo^{\ast}E|}

\newcommand{\evalk}{\eval_\Kcat}

\newcommand{\curryk}{\curry_\Kcat}
\newcommand{\prok}{\pro_\Kcat}
\newcommand{\proik}{\pi_{1\Kcat}}
\newcommand{\protk}{\pi_{2\Kcat}}

\SpecialCoor

\newcommand{\noderadius}{0.1}
\newcommand{\labelradius}{0.12}
\newcommand{\nodeline}{1pt}

\newcommand{\N}{90}

\newcommand{\W}{180}
\newcommand{\SWW}{202.5}
\newcommand{\SW}{225}

\renewcommand{\S}{270}

\newcommand{\SE}{315}
\newcommand{\SEE}{337.5}
\newcommand{\E}{0}
\newcommand{\NEE}{22.5}
\newcommand{\NE}{45}

\newcommand{\firsto}[3] 
{\cnode[linewidth=\nodeline](0,0){\noderadius}{#1}%
\uput{\labelradius}[#2](#1){$#3$}}

\newcommand{\firston}[3] 
{\cnode*[linewidth=\nodeline](0,0){\noderadius}{#1}%
\uput{\labelradius}[#2](#1){$#3$}}

\newcommand{\nexto}[6]
{\cnode[linewidth=\nodeline]([nodesep=#4,angle=#3]#1){\noderadius}{#2}%
\uput{\labelradius}[#5](#2){$#6$}}

\newcommand{\nexton}[6]
{\cnode*[linewidth=\nodeline]([nodesep=#4,angle=#3]#1){\noderadius}{#2}%
\uput{\labelradius}[#5](#2){$#6$}}

\newcommand{\sline}[2]
{\ncline[nodesep=0pt]{-}{#1}{#2}} 

\newcommand{\dline}[2]
{\ncline[nodesep=0pt,linestyle=dotted]{-}{#1}{#2}} 
 
\newcommand{\slineo}[6]
{\nexto{#1}{#2}{#3}{#4}{#5}{#6}%
\sline{#1}{#2}}

\newcommand{\slineon}[6]
{\nexton{#1}{#2}{#3}{#4}{#5}{#6}%
\sline{#1}{#2}}

\newcommand{\dlineo}[6]
{\nexto{#1}{#2}{#3}{#4}{#5}{#6}%
\dline{#1}{#2}}

\newcommand{\dlineon}[6]
{\nexton{#1}{#2}{#3}{#4}{#5}{#6}%
\dline{#1}{#2}}

\newcommand{\firstn}[2]
{\rput(0,0){\rnode{#1}{$#2$}}}
\newcommand{\nextn}[5]
{\rput([nodesep=#4,angle=#3]#1){\rnode{#2}{$#5$}}}

%% file: T1.tex
\keywords{domain, partial order, dcpo, natural domain, natural lub, typed lambda calculus, PCF, denotational semantics,
fully abstract model, non-cpo model, f-model,
game semantics 
 }
\subjclass{F.3.2, F.4.1}
\begin{abstract}
Normann proved that the domains of the game model of PCF
(the domains of sequential functionals) need not be dcpos.
Sazonov has defined natural domains for a theory of such incomplete domains.

This paper further develops that theory.
It defines lub-rules that infer natural lubs from existing natural lubs,
and lub-rule classes that describe axiom systems like that of natural domains.
There is a canonical proper subcategory of the natural domains,
the closed directed lub partial orders (cdlubpo),
that corresponds to the complete lub-rule class of all valid lub-rules.
Cdlubpos can be completed to restricted dcpos, which are dcpos that retain
the data of the incomplete cdlubpo as a subset.
\end{abstract}

\maketitle

\section{Introduction}

Is the tacit agreement (perhaps a kind of ``dogma'') of the directed completeness of semantic domains falling?
We might get this impression from the recent results of Dag Normann and Vladimir Sazonov
on the game model of PCF.
So we begin this introduction with a brief history of the models of PCF.

\subsection*{History}

PCF is a simply typed $\la$-calculus on integers with higher-order recursion.
The concept of PCF was formed by Dana Scott in 1969, see the historical document
\cite{Scott}.
It is used as a prototypical programming language to explore the relationship
between operational and denotational semantics, see the seminal paper of
Gordon Plotkin \cite{Plotkin}.
The first model of Scott was made of directed complete partial orders,
beginning with flat domains for integers and booleans and the full domain of continuous functionals
for higher-order types.
It was natural to demand directed completeness of the domains,
so that we get a definition of continuity that leads to closure under function spaces
and the existence of
all conceivable fixpoints for the semantics of recursive functions.
But the model contained ideal elements that were not realized in the language:
These were finite elements, like the ``parallel or'' function,
or infinite elements (as lubs of directed sets of finite elements).
The presence of 
the finite unrealized elements (like ``parallel or'') already causes the model to be not
fully abstract (\ie the denotational semantics does not match the operational one),
as was observed by Gordon Plotkin \cite{Plotkin}.

The question of a fully abstract model arose, where a model was supposed to consist of
directed complete partial orders,
following the established ``dogma''.
The first fully abstract cpo model was constructed by Robin Milner \cite{Milner} in 1977
from equivalence classes of finite combinator terms by an inverse limit of domains.
Later the same model was constructed by G\'erard Berry \Berry\ from equivalence classes
of proper PCF-terms by an ideal completion.

So this model was built on syntactic terms of the language, which was not considered satisfactory,
and the search for a purely mathematical fully abstract cpo model began.
The widely accepted solution was the game semantics after 1990
\cite{Abramsky/Jagadeesan,Hyland/Ong,Nickau}.
In game semantics a term of PCF is modeled by a strategy of a game,
\ie by a process that performs a dialogue of questions and answers with the environment, the opponent.
These strategies are still intensional; the \famm\ is formed by a quotient, the extensional
collapse.
The strategies can be identified with
PCF B\"ohm trees of a certain normal form,
see \cite[section 6.6]{Amadio/Curien}.

It was an open problem whether the model of game domains is isomorphic to Milner's fully abstract cpo-model,
\ie whether its domains are cpos and so contain every element of the cpo-model.
This problem was solved by Dag Normann \cite{Normann}: 
its domains are not cpos, \ie there are directed sets that have no lub.
The example given by Normann is in type 3 and rather sophisticated.
Then Vladimir Sazonov made a first attempt to build a general theory for these
non-cpo domains \cite{Sazonov:models,Sazonov:natural}.
His main important insight was that functions are continuous only with respect to certain lubs of directed sets
that he calls ``natural lubs''; these are the hereditarily pointwise lubs in PCF.
He defines an abstract structure of ``natural domains'' \cite{Sazonov:natural} as a partial order
with an operator that designates certain lubs of (general, not only directed)
subsets as ``natural lubs'', fulfilling some axioms.
He shows that the category of natural domains and functions that are continuous \wrt the
directed natural lubs is a cartesian closed category (ccc).
He defines naturally finite elements and natural algebraicity \wrt the natural directed lubs,
and also shows that naturally algebraic, bounded complete natural domains form a ccc.
In a recent paper Normann and Sazonov \cite{Normann/Sazonov} show that in the game model of PCF,
the sequential functionals,
there is a Normann-example in a second-order type, there
 are directed lubs that are not natural, and there are naturally finite elements 
that are not finite in the classical sense.
The main results of the last paper are also covered in the recent textbook
of Longley and Normann: ``Higher-order computability'' \cite{Longley/Normann},
section 7.6.

Generally speaking, all these problems are due to a fundamental mismatch between
the two worlds that semantics is relating:
The world of syntax, of mechanism (in the form of programming languages and abstract machines),
of intension on one side,
and the more abstract world of domains and continuous functions, of extension
on the other side.
The problems are generally caused by restrictions on the syntactic side.
So the restriction to sequentiality caused the full abstraction problem for PCF.
Its solution, games, are constructions that stand somewhat in the middle between the two worlds.

Sazonov's natural domain theory accounts for another syntactic restriction:
that limits of ascending chains of finite elements can only be formed for those
chains that are ascending with the syntactic order, as a Boehm tree.
(This is not bound to sequentiality, as Sazonov shows in \cite{Sazonov:models}
a corresponding model for PCF with parallel conditional.)
Natural domains model this incompleteness,
but they miss an important property:
the existence of fixpoints of endofunctions (on domains with $\bot$).
(The domains of game semantics have this property,
a mechanism always has the fixpoints by construction.)
So a category of abstract incomplete domains (\eg natural domains)
must be understood as a (cartesian closed) ``house'' in which several
more syntactic programming language models (with the fixpoint property) live together.
We pose as an open problem to find categories of abstract incomplete domains
with the existence of fixpoints.

We have seen that the game domain model, the model of sequential functionals,
results from a syntactic restriction to Boehm trees.
There are other syntactic restrictions conceivable,
the most extreme being a restriction just to the terms of PCF itself.
So in my paper \cite[section 3]{Mueller:berry} on Berry's conjectures
I have given the definition of a whole spectrum of
fully abstract models of PCF (``f-models'') as sets of ideals of equivalence classes of finite terms,
such that application is defined and every PCF-term has a denotation.
In this spectrum Milner's cpo model is the largest model,
the pure term model is the least,
and the game model is properly between the two.

\subsection*{Ideas of this paper}

In this paper we further explore the abstract domain theory of incomplete domains.
Our main objective is to find cartesian closed categories.
We begin high above the natural domains with the most general conceivable structure,
the directed-lub partial orders (dlubpo),
partial orders with designated directed lubs,
in the form of a relation $A\nlubD a$,
meaning that the directed subset $A$ has the natural lub $a$.
The only axiom they must obey is the singleton axiom $\{a\}\nlubD a$ 
(Sazonov's axiom 3).

We define lub-rule classes to classify axiom systems with a form like that
of natural domains.
A lub-rule on a partial order $D$ is a triple $(D,P\ruleto A)$
with $P$ a set of subsets of $\Del$ that each have a lub
and $A$ a subset of $\Del$ that has a lub.
This expresses the fact that in $D$ we can infer from the existence of lubs of the
elements of $P$ the existence of the lub of $A$.
This inference, this lub-rule, is ``valid'' if it is invariant under monotonic functions,
\ie every monotonic function $\fDE$ for some partial order $E$
that respects the lubs of the elements of $P$ respects also the lub of $A$.
We explore axiom systems (of dlubpos) that can be described by classes of valid lub-rules.

We explain the philosophical significance of our translation from an axiom system $S$ to
a lub-rule class as a ``partial extensionalization'' of (the intension of) $S$,
\ie an extension that is between the pure syntax of $S$ and the full extension,
the class of dlubpos that fulfill $S$.

The validity of a lub-rule can be characterized by a closure operator $\clD$ on
subsets of dlubpos $D$ that infers from one element all elements below it,
and from a natural subset the natural lub of it.
This closure operator already appeared in the work of Bruno Courcelle and Jean-Claude Raoult 
on completions of ordered magmas \cite{Courcelle/Raoult}.

A lub-rule class is complete if it encompasses all valid lub-rules.
The dlubpos generated by a complete lub-rule class are called closed dlubpos (cdlubpo).
They can be characterized by the closure operator $\clD$,
\ie they fulfill the axiom S9 (closure):\\
If $A\sub \Del$ is directed with lub $a$ and $a\in \clD A$, then $A\nlubD a$.\\
They form a ccc.
There are natural domains that are no cdlubpo.

In contrast to natural domains, cdlubpos have several characterizations
as canonical structures.
Cdlubpos are the dlubpos that are ``realized''
by ``restricted dcpos'' (rdcpo).
So complete domains are coming in again, and the ``dogma'' of completeness could be ``saved''.
The idea is to complete every cdlubpo with new improper elements
(the ``blind realizers'') to a dcpo that retains the data of the incomplete cdlubpo
as a subset, by an order embedding.

This idea of realization of a partial order by a dcpo goes back to Alex Simpson \cite{Simpson}:
In Simpson's approach every element of the partial order is realized by one or several realizers of the dcpo,
while every realizer of the dcpo realizes exactly one element of the partial order.
In our approach every element of the partial order is realized by exactly one realizer of the dcpo,
while every realizer of the dcpo realizes at most one element of the partial order.
(The two approaches could be combined, see the last section \ref{s:outlook} Outlook.)

In a sequel paper we will work out the connection between the categories 
$\Dlubpo$ and $\Rdcpo$.
There is an adjunction between them,
which establishes an adjoint equivalence between the sub-cccs 
$\Cdlubpo$ and $\Crdcpo$ (closed rdcpos).
This connection also makes it possible to transfer the theory of cccs of algebraic dcpos to the
realm of closed rdcpos \resp cdlubpos.

\subsection*{Outline of the paper}

\begin{enumerate}[1.]
\setcounter{enumi}{1}
\item Preliminaries and notation:\\
We repeat a concrete definition of cartesian closed categories
and basic definitions of dcpo theory.
Throughout the paper we will encounter closure and completion procedures
that all follow the same abstract scheme.
Here we extract this scheme as a fixpoint lemma on powersets and the definition
of ``rule systems'' with their deductions.
\item Sazonov's definition of natural domains revisited:\\
We repeat Sazonov's definition of natural domains,
and give a simpler equivalent axiom system.
Although we deviate from natural domains in the following sections,
we will refer to the new axioms that are inferred here.
We also show an example of a natural domain where natural lubs are
needed for general subsets, not only directed ones.
\item Directed-lub partial orders (dlubpo):\\
We define directed-lub partial orders as the most general structure we conceive.
In this category the exponents, if they exist, have a definition that
is generally different from the pointwise exponents of natural domains.
We give sufficient conditions on subcategories of $\Dlubpo$ to have
terminal, products and exponents like the normal ones.
The main theorems are that $\Dlubpo$ is no ccc,
but that the full subcategory of dlubpos that fulfill Sazonov's axiom S5
is already a ccc (with exponents of the general form).

This section contains the basic definitions of dlubpos,
but the results are mainly unrelated to the rest of the paper,
only few minor ones are used in the following sections.
For a first reading, I recommend to read only the basic definitions 4.1 to 4.4
and skip the rest of the section.
\item Lub-rule classes and closed dlubpos:\\
We introduce the (valid) lub-rules, lub-rule systems 
and (complete, invariant) lub-rule classes we described above.
We define closed dlubpos (cdlubpo) fulfilling the closure axiom S9.
These are the dlubpos fulfilling all valid (directed) lub-rules.
Every cdlubpo is a natural domain.
The dlubpos generated by an invariant lub-rule class form a full reflective
subcategory of $\Dlubpo$.
\item The ccc of S10-dlubpos:\\
We introduce a new axiom S10 for dlubpos.
The category of these dlubpos is the largest full sub-ccc of $\Dlubpo$
that is generated by an invariant lub-rule class and has the 
pointwise exponents.
Every natural domain and every cdlubpo is in this category.
\item Example of a natural domain that is no cdlubpo:\\
We first show by a simple finite example that the lub-rule class
corresponding to the axioms of natural domains is not complete.
Then we give an example of a natural domain that is no cdlubpo.
\item Algebraic dlubpos:\\
We show that every algebraic dlubpo that fulfills axiom S6 (cofinality) is a cdlubpo.
This means that algebraic natural domains and algebraic cdlubpos are the same.
\item Restricted partial orders and restricted dcpos:\\
We give a sufficient condition based on subcategory morphisms 
for a dlubpo to be a cdlubpo.
We define restricted partial orders (rpo) and restricted dcpos (rdcpo)
and show that cdlubpos are exactly the dlubpos realized by rpos \resp rdcpos.
\item
Outlook
\end{enumerate}

%% file: T2.tex
\section{Preliminaries and notation}

Notation:
We mostly write function application without brackets, if possible.
Function application associates to the left.
If $f$ is a function that is defined on the elements of a set $A$,
then we write $\fsetf A=\set{fa}{a\in A}$.
$\Pot(S)$ is the powerset of the set $S$.

We give some basic definitions of category theory and domain theory,
and then a fixpoint lemma on powersets with a definition of ``rule systems''.

\subsection{Category theory}

The main property of our categories of domains that interests us here is cartesian closedness.
We adopt a concrete definition from \cite[def. 4.2.5]{Amadio/Curien}:
\begin{defi}[cartesian closed category] Let $\Kcat$ be a category.\hfill
\begin{enumerate}[\em(1)]
\item
A \defin{terminal object} in $\Kcat$ is a $\terminal\in\Kcat$ such that:\\
$\all{C\in\Kcat}\exiun{h\typ C\fun \terminal}$
\item
A \defin{product} of $D,E\in \Kcat$ is an object $D\pro E\in \Kcat$
with \defin{projections} $\proi\typ D\pro E\fun D$
and $\prot\typ D\pro E\fun E$
such that:\\
$\all{C\in\Kcat}\all{f\typ C\fun D}\all{g\typ C\fun E}\exiun{h\typ C\fun D\pro E}
(\proi\comp h=f \text{ and } \prot\comp h=g).$\\
The morphism $h$ is denoted by $\lpro f,g\rpro$, $\lpro\_,\_\rpro$
is called the \defin{pairing operator}.\\
For $f\typ D\fun D'$, $g\typ E\fun E'$:
$f\pro g\typ D\pro E\fun D'\pro E'$ is defined as $f\pro g=\lpro f\comp\proi, g\comp\prot\rpro$.
\item
Let $\Kcat$ have all (binary) products.\\
An \defin{exponent} of $D,E\in\Kcat$ is an object $D\expo E\in \Kcat$
with an \defin{evaluation morphism} $\eval\typ (D\expo E)\pro D\fun E$ such that:\\
$\all{C\in\Kcat}\all{f\typ C\pro D\fun E}\exiun{h\typ C\fun(D\expo E)}
(\eval\comp(h\pro \id)=f).$\\
The morphism $h$ is denoted by $\curry(f)$,
$\curry$ is called the \defin{currying operator}.
\end{enumerate}
A \defin{cartesian closed category (ccc)} is a category that has a terminal object and 
all (binary) products and all exponents.
\end{defi}

We will also see adjunctions and reflective subcategories,
we will mainly use the notations of \cite{Maclane}.

\subsection{Domain theory}

We take our definitions from the excellent textbook \cite{Amadio/Curien}.

A structure $D=(\Del,\lexD)$ is a \defin{partial order (po)}
if $\Del$ is a set and $\lexD$ is a binary relation on $\Del$ 
that is reflexive, transitive and antisymmetric.\\
$\lexD$ is simply written $\lex$ if $D$ is clear from the context,
and this abbreviation also applies to other structures and indices.\\
A non-empty subset $A\sub\Del$ is \defin{directed}
if for all $a,b\in A$ there is some $c\in A$ with $a\lex c$ and $b\lex c$.\\
For a partial order $D$ (or some extension of a partial order with more data)
$\Potlub(D)$ is the set of subsets of $\Del$ that have a least upper bound (lub) in $D$;
and $\Potdlub(D)$ is the set of directed subsets of $\Del$ that have a lub in $D$.\\
For subsets $A,B\sub\Del$ it is written $A\lset B$ if 
for all $a\in A$ there is $b\in B$ with $a\lex b$;
and for $A\sub \Del$, $b\in \Del$: $A\lex b$ if for all $a\in A$ it is $a\lex b$.

$D$ is a \defin{directed complete partial order (dcpo)}
if every directed subset of $A\sub\Del$ has a least upper bound (lub), denoted $\Lubex A$.
If futhermore $D$ has a least element (written $\bot$),
then it is called a \defin{complete partial order (cpo)}.\\
For $D,E$ dcpos, a function $f\typ\Del\fun\Eel$ is \defin{continuous} 
if it 
preserves directed lubs: for all directed $A\sub \Del$ it is $f(\Lubex A)=\Lubex(\fsetf A)$.
(Then it is also monotonic,
\ie $a\lexD b \ifthen fa\lexE fb$.)

$\Dcpo$ is the category of dcpos and continuous functions.
The continuity of $f$ is specified by $f\typ D\fun E$,
which always means that $f$ is a morphism in the category of $D$ and $E$.\\
$\Dcpo$ is a ccc:
The terminal object is the one point dcpo.
For dcpos $D,E$ the product is $D\pro E= (\Del\pro\Eel, \lex_{D\pro E})$
where $\lex_{D\pro E}$ is pointwise.
The exponent is $D\expo E = (\elem{D\expo E},\lex_{D\expo E})$
where $\elem{D\expo E}$ is the set of continuous functions,
and $\lex_{D\expo E}$ is the pointwise order.
Lubs of directed sets of continuous functions are taken pointwise.

\subsection{Fixpoint lemma and rule systems}

We will encounter many closure and completion procedures that all follow a certain abstract
scheme that is here extracted.
These results are certainly all well known,
but I did not find a reference with proofs in the literature that matches.
We will use ordinals, a short introduction to them can be found in \cite{Johnstone}.

\begin{defi}
Let $S$ be a set.\\
A function $f\typ \Pot(S)\fun \Pot(S)$ is a \defin{\preclosure} 
if it is \defin{increasing}: for all $A\sub S$ it is $fA\sups A$,
and it is \defin{monotonic}: for $A,B\sub S$, $A\sub B$ implies $fA\sub fB$.\\
An $A\sub S$ with $fA=A$ is called \defin{closed under} $f$.\\
For $A\sub S$ we define $f^0 A=A$, $f^{\alpha+1}A=f(f^{\alpha}A)$ for all ordinals $\alpha$,
$f^{\beta}A = \Join_{\alpha<\beta} f^{\alpha}A$ for all limit ordinals $\beta$.
\end{defi}

\begin{lem}[fixpoint lemma]\hfill\\
Let $S$ be a set, $f\typ \Pot(S)\fun\Pot(S)$ a \preclosure\ and $A\sub S$.\\
Let $B$ be the intersection of all $A'\sups A$ with $fA'=A'$.\\
Then $B$ is the least set with $A\sub B\sub S$ and $fB=B$.
$B$ is called the \defin{closure} of $A$ under $f$.\\
The map from $A\sub S$ to the closure of $A$ under $f$ is called 
the \defin{closure operator} for $f$.\\
It is $B=\Join_{\alpha\text{ ordinal}} f^{\alpha}A$.
Furthermore $B=f^{\gamma}A$ for some ordinal $\gamma$.
\end{lem}
\proof
It is $fB\sups B$ because $f$ is increasing.\\
It is $fB\sub fA'\sub A'$ for all $A'\sups A$ with $fA'=A'$, therefore $fB\sub B$.\\
Let $C=\Join_{\alpha \text{ ordinal}}f^{\alpha}A$.
We show $C\sub B$, \ie $f^\alpha A\sub B$ for all ordinals $\alpha$, by induction on $\alpha$:\\
It is $f^0 A=A\sub B$.
$f^{\alpha+1}A=f(f^\alpha A)\sub fB=B$.\\
$f^{\beta}A= \Join_{\alpha<\beta}f^\alpha A\sub B$ for all limit ordinals $\beta$.\\
To prove $B\sub C$ we show that $fC=C$:\\
By Hartogs' lemma \cite{Hartogs}\cite[lemma 7.1]{Johnstone}, there is an ordinal $\gamma$
that cannot be mapped injectively into $S$.
Assume $f(f^\gamma A) \neq f^\gamma A$.\\
Then for every element $\alpha$ of $\gamma$, \ie for every ordinal $\alpha<\gamma$,
we have $f(f^\alpha A)\neq f^\alpha A$.\\
If we map each $\alpha<\gamma$ to some new element of $f(f^\alpha A)$ that is not in $f^\alpha A$,
we get an injective map from $\gamma$ into $S$, contradiction.
So it must be $f(f^\gamma A)=f^\gamma A=C$.\\
As $B$ is the least set with $fB=B$ and $A\sub B$, we get $B\sub C$.
\qed

In all our applications of the fixpoint lemma,
the \preclosure\ is generated by a rule system on the underlying set $S$:

\begin{defi}[rule system]\label{d:rulesystem}
A \defin{rule system} on a set $S$ is a relation $\ruleto\,\sub \Pot(S)\pro S$.\\
The elements of $\ruleto$ are called \defin{rules}.
The \defin{\preclosure\ of} this rule system is\\
$f\typ \Pot(S)\fun\Pot(S)$ defined by 
$fA= A\join \set{a\in S}{\exi{B\sub A} B\ruleto a}$.\\
The closure operator for $f$ is called the \defin{closure operator} of the rule system,
corresponding to \defin{closures under} the rule system.\\
A \defin{deduction of} $a\in S$ \defin{from} $A\sub S$ in this rule system is some $(N,r,\lab,\pre)$
where $N$ is the set of \defin{nodes},
$r\in N$ is the \defin{root},
$\lab\typ N\fun S$ is the \defin{labelling function},\\
$\pre\typ \set{n\in N}{\lab n\notin A}\fun \Pot(N)$ is the \defin{predecessor function},\\
such that $\lab r=a$,\\
for all $n\in N$ with $\lab n \notin A$: $\fset{\lab}(\pre n)\ruleto \lab n$,\\
(we say that $\lab n$ is \defin{deduced} from $\fset{\lab}(\pre n)$),\\
for all $n\in N$ there is a unique finite path of length $i\geq 1$:\\
$r=n_1$, $n_2\in \pre n_1$, $n_3\in \pre n_2$,\ldots, $n_i=n$,\\
and for all $n\in N$ there is no infinite path $n=n_1$, $n_2\in \pre n_1$, $n_3\in \pre n_2$,\ldots.
(So such paths end in some $n_i\in N$ with $\lab n_i\in A$.)

If there is a deduction of $a$ from $A$,
then the pair $(A,a)$ is called a \defin{derived rule} of the rule system.
\end{defi}

So the deduction is a tree with root $r$ labelled $a$ and leaves labelled by elements of $A$,
such that each non-leaf node is deduced by a rule from its predecessors.
It has a well-founded structure and we can prove properties of its nodes by induction on this
structure.

\begin{lem}
Let $\ruleto$ be a rule system on a set $S$, and $f$ its \preclosure.
Let $A\sub S$ and $a\in S$.\\
There is a deduction of $a$ from $A$ in $\ruleto\ \ifff a$ is in the closure of $A$ under $f$.
\end{lem}
\proof
$\ifthen$:
Let $(N,r,\lab,\pre)$ be the deduction of $a$ from $A$ in $\ruleto$.
We prove by induction on the deduction that for every node $n\in N$,
$\lab n$ is in the closure $B$ of $A$ under $f$.
This is clear as $A\sub B$ and $fB=B$.

$\thenif$:
We show by induction on $\alpha$ that for every ordinal $\alpha$
and every $a\in f^{\alpha} A$,
there is a deduction of $a$ from $A$.
It is clear for $\alpha=0$ and $\alpha$ a limit ordinal.
For $\alpha=\beta+1$,
$a$ is already in $f^{\beta}A$ or else there is some $C\sub f^\beta A$ with $C\ruleto a$.
In the second case, we build the deduction with root $r$ and $\lab r=a$
and connect $r$ to all the deductions of the elements of $C$ as predecessors.
\qed

%% file: T3.tex
\section{Sazonov's definition of natural domains revisited}\label{s:sazonov}

In this section we repeat Sazonov's definition of natural domains, and
give a simpler equivalent axiom system.
We also give an example that shows
that natural lubs are needed for general subsets,
not only directed ones.

Sazonov's natural domains \cite{Sazonov:natural} are a general extrapolation of
the domains of the game model of PCF.
They are just partial orders with a designation of some of the (general) subsets
with lubs as ``natural lubs''.
The natural lubs of directed sets must be respected by the naturally continuous functions.
The natural lubs have to obey certain closure conditions which are
taylored to ensure cartesian closedness of the resulting category under a certain
function space.

We translate the definition into a new notation:
instead of the partial natural-lub-operator $\sazlub$ we use a relation
symbol $\nlub$ for  natural convergence.

\begin{defi}[Sazonov, def. 2.1 of \cite{Sazonov:natural}] $\phantom{X}$\\
A structure $D=(\Del,\lexD,\nlubD)$ is a \defin{natural domain} if\\
$\Del$ is a set of \defin{elements},\\
$\lexD$ is a partial order on $\Del$,\\
$\nlubD\; \sub \Potdlub(\Del)\pro \Del$ is the \defin{(natural) convergence relation} on directed sets,\\
($X\nlubD x$ means: $X$ has the \defin{natural lub} $x$,
such an $X$ is called \defin{natural})\\
such that $\nlubD$ can be extended by pairs $(X,x)$
with $X\sub\Del$ \emph{not} directed, $x\in \Del$, to a relation on all subsets,
which we also denote $\nlubD$,
and the following axioms are fulfilled  for the extended relation
(where axiom Sx corresponds to $\sazlub$x in Sazonov):\\
\textbf{(S1)} If $X\nlub x$, then $\Lubex X = x$.\\
\textbf{(S2)} If $X\sub Y \sub \Del$, $X\nlub x$ and $Y\lex x$, then $Y\nlub x$.\\
\textbf{(S3 singleton)} For all $x\in \Del$: $\{x\}\nlub x$.\\
\textbf{(S4(1))} If $\yijIJ$ is a non-empty two-parametric family of elements in $\Del$,\\
and for every $i\in I$ it is $\yijJ \nlub d_i$,\\
then $\{d_i\}_{i\in I} \nlub d \ifff \yijIJ\nlub d$.\\
\textbf{(S4(2))} If $I$ is an index set with a directed partial order, and $\yijII$ is a two-parametric family
of elements in $\Del$ that is monotonic in each parameter $i$ and $j$,\\
then $\yijII \nlub y  \ifff \yiiI \nlub y$.
\end{defi}
Note that we deviate from Sazonov's definition in that we take only
the part of the convergence relation on \emph{directed} subsets as the data of the structure,
and leave the existence of the extended relation as a side condition.
This is justified by the morphisms of the category,
the (naturally) \cont\ functions,
which must be \cont\ only \wrt the \emph{directed} natural lubs.
We get our natural domains by identifying Sazonov's natural domains
which have the same order and the same \emph{directed} natural lubs.
The result is an equivalent category.
We make this change because we define all other related structures of
incomplete domains by convergence of directed sets only.

The specification of the lub in the conclusions of the axioms is redundant,
it suffices to specify that the set in question is natural,
\eg in axiom (S2): $Y$ is natural instead of $Y\nlub x$.

We do not count S1 as a proper axiom,
instead we treat it as a condition for the relation $\nlub$ 
that is always presupposed in the sequel.
The other proper axioms have a different character,
they deduce the naturalness of lubs from existing natural lubs.

We have translated the axiom system into a form 
that clearly discerns its different parts.
It needs a clean-up and simplification.
First, we can further split axiom S4(1) into the two axioms
S4(1\Right) and S4(1\Left),
where the logical equivalence $\ifff$ is replaced by the indicated direction.

\begin{prop}
In the presence of axiom S3 (singleton):
axioms S2 and S4(1\Left) together are equivalent to the axiom:\\
\textup{\textbf{(S6 cofinality)}} If $X,Y\sub \Del$, $X\nlub x$ and $X\lset Y\lset x$, then $Y\nlub x$.
\end{prop}
\proof
S2 and S4(1\Left) $\ifthen$ S6:\\
Let $X\nlub x$ and $X\lex Y\lex x$.\\
Let $I=\set{(a,b)}{a\in X, b\in Y\text{ and }a\lex b}$ and $J=\{1,2\}$.\\
For $(a,b)\in I$ define $y_{(a,b)1}=a$ and $y_{(a,b)2}=b$ and $d_{(a,b)}=b$.\\
For every $(a,b)\in I$ it is $\{y_{(a,b)j}\}_{j\in J} \nlub d_{(a,b)}$, by axiom S3 and S2.\\
It is $\{y_{ij}\}_{i\in I,j\in J} \nlub x$ by axiom S2,
as $X\sub \yijIJ$ by $X\lset Y$.\\
We apply axiom S4(1\Left) and deduce $\{d_{(a,b)}\}_{(a,b)\in I} \nlub x$.\\
This set is a subset of $Y$, therefore by axiom S2 we get $Y\nlub x$.\\
The reverse direction is immediate.
\qed

\begin{prop}
Axiom S4(2) is redundant, it follows from axiom S6 (cofinality).
\qed
\end{prop}

Sazonov proposes an ``optional clause, which might be postulated as well'',
as a replacement for axiom S4(2):\\[\med]
\textbf{(S5)} If $X\sub Y\sub \Del$, $Y\nlub y$ and $Y\lset X$, then $X\nlub y$.\\[\med]
This does not lead to a stronger axiom system:

\begin{prop}
Axioms S2 and S5 together are equivalent to axiom S6 (cofinality).
\end{prop}
\proof
S2 and S5 $\ifthen$ S6:\\
It is $X\sub X\join Y\sub \Del$, $X\nlub x$, $X\join Y\lex x$, so $X\join Y\nlub x$ by axiom S2.\\
It is $Y\sub X\join Y\sub\Del$, $X\join Y\nlub x$, $X\join Y\lset Y$, so $Y\nlub x$ by axiom S5.\\
The reverse direction is immediate.
\qed

Next, we want to give axiom S4(1\Right) in a form with sets instead of two-parametric families,
to get rid of the index sets of unlimited size:

\begin{prop}
In the presence of axioms S3 (singleton) and S6 (cofinality), axiom S4(1\Right) is equivalent to the axiom:\\
\textup{\textbf{(S7 transitivity)}}
If $\Xb\sub \Pot(\Del)$ with $X\nlub d_X$ for all $X\in \Xb$,\\
then $\set{d_X}{X\in \Xb} \nlub d  \ifthen  \Join \Xb \nlub d$.
\end{prop}
\proof
S4(1\Right) follows immediately from S7.\\
For the reverse direction, we prove S7 from S4(1\Right).\\
First, we have to prove the two special cases (1) $\Xb = \emptyset$ and (2) $\Xb=\{\emptyset\}$.\\
(1) The conclusion of S7 reads $\emptyset\nlub d \ifthen \emptyset\nlub d$.\\
(2) It is $\emptyset\nlub d_{\emptyset}$ and the conclusion reads $\{d_{\emptyset}\}\nlub d \ifthen
\emptyset\nlub d$.\\
Now assume we do not have case (1) or (2).\\
Take $I=\Xb$ and $J=\Join\Xb$.\\
Both $I$ and $J$ are not empty, as is required by S4(1\Right).  We define $\yij$:\\
If $i\in I=\Xb$ is not empty, 
then for all $j\in J$ take $\yij=j$ if $j\in i$ and else some arbitrary element of $i$.\\
If $i\in I$ is empty, then for all $j\in J$ take $\yij = \Lubex \emptyset$ (the least element),
this exists as $i=\emptyset \nlub d_{\emptyset} = \Lubex \emptyset$.\\
Then for all nonempty $i\in I$ it is $i\sub\yijJ$ and $\yijJ\lset d_i$,\\
so by axiom S6 it is $\yijJ\nlub d_i$.\\
If $i\in I$ is empty,
then it is $\yijJ = \{\Lubex \emptyset\}\nlub \Lubex \emptyset$ by axiom S3.\\
So in every case it is $\yijJ\nlub d_i$,
and we can draw the conclusion of S4(1\Right).\\
Furthermore, if not $\emptyset \in I$,
then $\yijIJ = \Join\Xb$, and the conclusion of S7 follows.\\
If $\emptyset\in I$, then $\yijIJ= \Join\Xb\join\{\Lubex\emptyset\}$.\\
From $\yijIJ\nlub d$ follows $\Join\Xb\nlub d$ by axiom S6.
This is the conclusion of S7.
\qed

In summary, we get a nice simpler axiom system as a base for the rest of the paper:
\begin{prop}\label{p:sazonov}
The axioms of the Sazonov natural domain are equivalent to
S1, S3 (singleton), S6 (cofinality) and S7 (transitivity).
\qed
\end{prop}

We can further combine axioms S6 and S7, for this we need a new definition:
\begin{defi}\label{d:under}
Let $D$ be the structure in the beginning of the definition of the Sazonov natural domain,
and $A,B\sub\Del$.\\
$A$ is \defin{under} $B$, $A\under_D B$, if for all $a\in A$:\\
there is $b\in B$ with $a\lexD b$ or
there is $B'\sub B$ with $B'\nlubD a$.
\end{defi}

\begin{prop}\label{p:under}
In the presence of axiom S3 (singleton):
axioms S6 (cofinality) and S7 (transitivity) together are equivalent to the axiom:\\
\textup{\textbf{(S8 under)}} 
If $A,B\sub\Del$, $A\nlub a$ and $A\under B \lset a$, then $B\nlub a$.
\end{prop}
\proof
S6 and S7 $\ifthen$ S8:\\
Let $\varphi\typ A\fun \Pot(\Del)$ be defined as:\\
$\varphi x =\{x\}$ if there is $y\in B$ with $x\lex y$,\\
otherwise take a choice of some $B'\sub B$ with $B'\nlub x$ and define $\varphi x = B'$.\\
Let $C=\Join(\fset{\varphi} A)$.
By the singleton axiom S3 and transitivity axiom S7 it is $C\nlub a$.\\
It is $C\lset B\lset a$, therefore $B\nlub a$ by the cofinality axiom S6.\\
The reverse direction is immediate.
\qed

So the axioms of the Sazonov natural domain are equivalent to S1, S3 and S8.\\
The proper axioms share a common form:
They are rules (in the sense of rule systems, definition \ref{d:rulesystem})
that deduce from the existence of the lubs of some sets
the existence of the lub of another set,
all in some configuration of a partial order.
One can get the impression that the  axioms are ``complete'' in this respect,
that all ``necessary'' deductions can be made.

In section \ref{s:lubrule} we set up the general frame of lub-rule systems
where we can define what this completeness means.
It will turn out in section \ref{s:incomplete} that the axioms are in fact incomplete.
So we will have to replace them by a stronger axiom in order to get
a class of domains that can be realized by dcpos.

We now come to the morphisms of the category of natural domains.
\begin{defi}[Sazonov def. 2.3(a) in \Saz] \hfill\\
Let $D,E$ be natural domains.\\
A function $f\typ \Del\fun\Eel$ is \defin{(naturally) continuous}
if it is monotonic \wrt $\lexD$ and $\lexE$,
and if $X$ is directed and $X\nlubD x$, then $\fset{f}X\nlubE fx$.
(Monotonicity is here redundant.)\\
If $f$ is naturally continuous, we write $f\typ D\fun E$.
\end{defi}

In \Saz\ it is shown that the natural domains with naturally continuous functions
form a ccc.\\
The terminal object is the one point domain $\{\{\bot\},\bot\lex \bot,\{\bot\}\nlub \bot\}$.\\
The product $D\pro E$ has $\elem{D\pro E}=\Del\pro\Eel$,
$\lex_{D\pro E}$ is component-wise,
and $A\nlub_{D\pro E} a$ iff $\fset{\proi} A\nlubD \proi a$ and $\fset{\prot} A\nlubE \prot a$.\\
The exponent $D\expo E$ has $\elem{D\expo E}$ the set of naturally continuous functions,
$\lex_{D\expo E}$ is pointwise,
and $F\nlub_{D\expo E} f$ iff $Fx\nlubE fx$ for all $x\in \Del$.

As we remarked above, we take only the part of the convergence relation on directed sets
as the data of a natural domain,
and leave the existence of the extended relation as a side condition.
The following example shows a case where natural lubs of general subsets are needed
as intermediary steps in the deduction of the natural lubs of directed subsets,
so that the extended relation is really needed.

\begin{figure}[htb]
\begin{pspicture}(-0.7,-0.2)(2.0,6.5)
\firsto{b11}{\W}{b_{11}}
\slineo{b11}{b12}{\N}{1}{\W}{b_{12}}
\dlineon{b12}{b1}{\N}{1}{\W}{b_1}
\nexto{b1}{b21}{\N}{1}{\W}{b_{21}}
\slineo{b21}{b22}{\N}{1}{\W}{b_{22}}
\dlineon{b22}{b2}{\N}{1}{\W}{b_2}
\slineo{b1}{a1}{\SEE}{1.5}{\E}{a_1}
\slineo{b2}{a2}{\SEE}{1.5}{\E}{a_2}
\sline{a1}{a2}
\dlineon{a2}{a}{\N}{1.5}{\E}{a}
\sline{b1}{a}
\sline{b2}{a}
\end{pspicture}
\caption{Natural domain $E$ (partial)}\label{f:weakE}
\end{figure}

Figure \ref{f:weakE} shows only part of the natural domain $E$.
It first has
the elements $a$, $a_i$ and $b_i$ for $i\geq 1$.
The order is $a_i\lex a_j$ iff $i\leq j$,
$a_i\lex a$, $a_i\lex b_i$ and $b_i\lex a$ for $i\geq 1$.
And $\aiset\nlub a$ is natural. 
It has further the elements of $B=\{b_{ij}\}_{i,j\geq 1}$ with ($b_{ij}\lex b_{ik}$
iff $j\leq k$) and $b_{ij}\lex b_i$ for all $i\geq 1$.
It is stipulated that the $\{b_{ij}\}_{j\geq 1}\nlub b_i$ are natural.
Moreover, we erect an ``artificial'' directed set over the elements of $B$,
whose elements are not depicted in the diagram:
every finite subset $X\sub B$ is an element of $E$,
the set of these elements is called $\Bb$.
For all $i,j$ it is defined $b_{ij}\lex \{b_{ij}\}$.
For $X,Y\in \Bb$ it is defined $X\lexE a$ and ($X\lexE Y$ iff $X\sub Y$).
$E$ is defined to be the natural domain that is completed by all conclusions of the
partial order axioms and the axioms of naturality.
$\Bb$ is a directed subset of $\Eel$ with the lub $a$.

From $\aiset\nlub a$ we deduce $\biset\nlub a$ by axiom S6,
then $B\nlub a$ by axiom S7,
then $\Bb\nlub a$ by axiom S6.

Here $\biset\nlub a$ and $B\nlub a$ are intermediary steps that deduce the naturality of non-directed subsets.
There is no deduction of the directed natural lub $\Bb\nlub a$
from the given directed natural lubs that does not use a
non-directed intermediary step.

%% file: T4.tex
\section{Directed-lub partial orders (dlubpo)}

Here we explore categories larger than the category of natural domains.
We want to find a ``minimal part'' of the natural domain axioms 
that already achieves cartesian closedness.
Less axioms means more chaos.
We always presuppose axiom S1, which we do not count as proper axiom.
When there is no other axiom valid,
we  get the normal terminal object $\terminal$,
but we do not get all constant element morphisms $\la x.d\typ \terminal \fun D$, $d\in D$
some object.
So in order to constrain chaos,
we start with the ``most obvious'' axiom, the singleton axiom S3,
and get the category $\Dlubpo$ of directed-lub partial orders.
These are partial orders with a designation of some lubs of \emph{directed} subsets as natural.
This category has the expected terminal object, products
and the constant element morphisms.
We prove that the exponents do not always exist;
and if they exist, they are generally defined differently from those in the category of
natural domains.
(But the natural domains,
with their \emph{directed} natural subsets as data, are a full sub-ccc of $\Dlubpo$ 
such that their exponents coincide with the general exponents of $\Dlubpo$.)
We also give sufficient conditions on subcategories of $\Dlubpo$
to have products and exponents like the normal ones,
in analogy to lemma 5 of \cite{Smyth}.
We show that the dlubpos fulfilling axiom S5 (for directed sets) already form a ccc.

For a first understanding of the following sections only definitions 4.1 to 4.4
are necessary.
The rest of this section may be skipped,
although definition \ref{d:genfun},
propositions \ref{p:evalstar} and \ref{p:expo1}(1) will be used in the following sections.


\begin{defi}
A structure $D=(\Del,\lexD,\nlubD)$ is a \defin{directed-lub partial order (dlubpo)} if\\
$\Del$ is a set of \defin{elements} of $D$,\\
$\lexD$ is a partial order on $\Del$, lubs are denoted by $\Lubex_D$,\\
$\nlubD\; \sub \Potdlub(\Del,\lexD)\pro\Del$ is the \defin{(natural) convergence relation},\\
if $A\nlubD a$, then $A$ is a directed subset of $\Del$,
and $a\in \Del$ is its lub,\\
($a$ is called the \defin{natural lub} of $A$, $A$ is called a \defin{natural} (directed) subset),\\
$\nlubD$ fulfills the \defin{singleton axiom} S3:
For every $d\in \Del$ it is $\{d\}\nlubD d$.
\end{defi}

\begin{defi}
Let $D,E$ be dlubpos.\\
A function $f\typ\Del\nlub\Eel$ is \defin{continuous} if it is monotonic \wrt $\lexD$ and $\lexE$
and if $X\nlubD x$, then $\fset{f}X\nlubE fx$.
If $f$ is continuous, we write $f\typ D\fun E$.\\
The dlubpos with continuous functions form a category $\Dlubpo$,
with normal function composition and identity functions.
\end{defi}

Note that Sazonov calls such a function ``naturally continuous'',
which we abbreviate to ``continuous'',
as it is clear from the context that $D,E$ are dlubpos.

\begin{defi}
The dlubpo $\terminal=(\{\bot\},\bot\lex\bot, \{\bot\}\nlub \bot)$
is called the \defin{terminal dlubpo}.\\
$\terminal$ is the categorical terminal object in $\Dlubpo$.
It need not be so in subcategories.
If the subcategory has $\terminal$ as terminal object,
we say that it has the normal terminal.
\end{defi}

\begin{defi}
Let $D,E$ be dlubpos.\\
We define the \defin{product dlubpo} $D\pro E = (\DproEel, \lex_{D\pro E}, \nlub_{D\pro E})$
as follows:\\
$\DproEel = \Del \pro \Eel$,\\
$(a,b)\lex_{D\pro E} (a',b')$ iff $a\lexD a'$ and $b\lexE b'$,\\
if $A\sub\DproEel$ is directed and $(a,b)\in \DproEel$,
then $A\nlub_{D\pro E} (a,b)$ iff $\fset{\proi} A \nlubD a$ and $\fset{\prot} A \nlubE b$,
where we have the normal \defin{projections}
$\proi\typ \DproEel\fun \Del$ and
$\prot\typ \DproEel\fun \Eel$.\\
(It is clear that $(a,b)$ is the lub of $A$.)\\
We also have the normal pairing functions:\\
For any dlubpo $C$, $f\typ C\fun D$ and $g\typ C\fun E$
it is $\lpro f,g\rpro\typ C\fun D\pro E$, $\lpro f,g\rpro x = (fx,gx)$.\\
All these functions are continuous,
and $D\pro E$ with $\proi,\prot$ is the categorical product in $\Dlubpo$.
It need not be so in subcategories.
If the subcategory has the $D\pro E$ as categorical products,
we say it has the normal products.
\end{defi}

\begin{defi}\label{d:genfun}
Let $\Kcat$ be a subcategory of $\Dlubpo$, and $D,E \in \Kcat$.\\
We define (polymorphically) the function 
$\eval\typ \set{f}{f\typ D\fun E \text{ in } \Kcat}\pro \Del\fun\Eel$ by $\eval(f,d)=fd$.\\
We define the \defin{general function dlubpo} $\DexpoKSE$ (relative to $\Kcat$) by:\\
$\DexpoKSEel = \set{f}{f\typ D\fun E \text{ in } \Kcat}$,\\
$f\lex_{\DexpoKSE} g$ iff for all $x\in \Del$ it is $fx\lexE gx$,\\
if $F\sub \DexpoKSEel$ is directed and $f\in \DexpoKSEel$, then
$F\nlub_{\DexpoKSE} f$ iff
for all $A\sub\DexpoKSEel\pro \Del$ that are directed \wrt the component-wise order
with $\fset{\proi} A=F$ and $\fset{\prot} A \nlub d$ it is $\fset{\eval} A\nlubE fd$.\\
For $\Kcat= \Dlubpo$ we define $(\DexpoSE) = (D\expo_{\Dlubpo}^{\ast} E)$.\\
We also have the curried functions:
For any dlubpo $C$, $f\typ C\pro D\fun E$ in $\Dlubpo$ we define
$\curry(f)\typ |C|\fun \DexpoSEel$ by $\curry(f)c= \la d.f(c,d)$.
\end{defi}

We have to prove that in the definition $\Lubex F =f$:
for all $d\in\Del$, take $A=F\pro\{d\}$,
then $\fset{\eval} A\nlub fd$,
so $f$ is the pointwise lub of $F$.\\
$\DexpoKSE$ fulfills the singleton axiom:
for $f\in \DexpoKSEel$ it is $\{f\}\nlub f$ because $f$ is continuous.

The function $\eval$ is continuous.
The trick of the definition is that not only the continuity in constant second arguments
(as in the case of natural domains),
but the whole continuity of $\eval$ is coded in the definition of $\nlub_{\DexpoKSE}$.
We will see that if $\Dlubpo$ contains an exponent of $D$ and $E$,
then it is isomorphic to $\DexpoSE$.

\begin{prop}\label{p:evalstar}
$\eval\typ(\DexpoKSE)\pro D\fun E$ is continuous.\\
The results of $\curry(f)$, $\curry(f)c$ for all $c\in |C|$, are continuous.\\
$\curry(f)$ is monotonic, but need not be continuous.
(We will see a counter-example below.)
\qed
\end{prop}

We will now give sufficient conditions for subcategories of $\Dlubpo$ to contain terminal objects,
products and exponents that coincide partially with the normal terminals, products
and the general function dlubpos.
These conditions are extracted from the corresponding results of \cite[lemma 5]{Smyth}
for full subcategories of the category of $\omega$-algebraic cpos.
Those results apply generally to partial order structures,
and they achieve full coincidence, \ie isomorphism,
only for the order structure; while in our case the isomorphism does not extend 
to the additional structure of natural convergence.
The natural convergence of a dlubpo can be chosen more arbitrarily.
The only way to achieve full isomorphism in our case seems to presuppose the normal structure 
in the subcategory (or a set of structures that all together force the normal structure 
on the categorical object in question),
which we will do afterwards.

Smyth's results are for full subcategories,
while we give them for general subcategories,
by extracting the needed morphisms from Smyth's proofs as conditions in the propositions.
The proofs are adaptations of Smyth's proofs.
We give the proofs mainly for those results that we need in the sequel.

\begin{prop}
Let $\Kcat$ be a subcategory of $\Dlubpo$.
If $\Kcat$ contains some object that is not empty,
and $\Kcat$ contains a terminal object $D$ and all constant morphisms $\la x.a\typ D\fun D$
for $a\in \Del$,
then $D$ is isomorphic to the terminal dlubpo $\terminal=\{\{\bot\},\bot\lex\bot,\{\bot\}\nlub\bot\}$.
\end{prop}
\proof
$D$ cannot have two different elements,
since this would entail two different constant morphisms $D\fun D$.
$D$ cannot be empty, as there is a non-empty object.
Thus $D$ has just one element $a$,
and it must be $\{a\}\nlub a$ by the singleton axiom.
\qed

\begin{prop}[Smyth, lemma 5(ii) in \cite{Smyth}, also prop.\ 5.2.17(2) in \cite{Amadio/Curien}]\hfill\\
Let $\Kcat$ be a subcategory of $\Dlubpo$ with the normal terminal object $\terminal$.\\
Let $D,E\in\Kcat$ and $D\prok E$ be their categorical product in $\Kcat$,
with the projections $\proik,\protk$ and pairing functions $\lpro f,g \rpro_\Kcat$.\\
Let $\Kcat$ have all constant functions $\la x.a\typ \terminal\fun D$, for $a\in \Del$,
and $\la x.a\typ \terminal\fun E$, for $a\in \Eel$.\\
(As $D,E$ fulfill the singleton axiom, all these functions are sure to be continuous,
so are morphisms in every \emph{full} subcategory $\Kcat$.)
\begin{enumerate}[(1)]
\item The function $\varphi\typ D \prok E\fun D\pro E$, $\varphi=\lpro\proik,\protk\rpro$,
is bijective.
(And of course it is continuous, so a morphism in $\Dlubpo$.)\\
The translation $\vphi$ respects projections and pairing functions:\\
$\proi\comp \vphi= \proik$,
$\prot\comp \vphi= \protk$,
$\vphi\comp\lpro f,g\rpro_{\Kcat}=\lpro f,g\rpro$.
\item
Let, furthermore, some $C\in \Kcat$ with some $c,c'\in\Cel$, $c\lex c'$ and $c\neq c'$,
and for all $d\lex d'$ in $D$ 
the function $f\typ C\fun D$ be in $\Kcat$ with $fx=d$ for $x\lex c$ and $fx=d'$ else,
and likewise for all $e\lex e'$ in $E$
the function $f\typ C\fun E$ with $fx=e$ for $x\lex c$ and $fx=e'$ else.\\
(As to the existence of these functions in \emph{full} subcategories, the remark above
applies again.)\\
Then $\vphi^{-1}$ is monotonic, so that $\vphi$ is an order-isomorphism.
$\vphi^{-1}$ is continuous in each component of its argument separately.
(But it need not be \cont.)
\end{enumerate}
\end{prop}
\proof is an adaptation of the proof of part (ii), lemma 5 of \cite{Smyth}.
\qed

\begin{prop}[Smyth, lemma 5(iii) in \cite{Smyth}, also prop.\ 5.2.17(3) in \cite{Amadio/Curien}]\label{p:expo1}
\hfill\\
Let $\Kcat$ be a subcategory of $\Dlubpo$ with the normal terminal object $\terminal$
and all normal products $D\pro E$.\\
Let $D,E\in \Kcat$ and $\DexpoKE$ be their categorical exponent in $\Kcat$,
with the evaluation\\
 $\evalk \typ (\DexpoKE)\pro D\fun E$
and curried morphisms $\curryk(f)$.\\
Let $\Kcat$ have all constant functions $\la x.f\typ \terminal\fun(\DexpoKE)$,
for $f\in \DexpoKE$.\\
(As $\DexpoKE$ fulfills the singleton axiom,
all these functions are sure to be \cont, so are morphisms in every \emph{full} subcategory $\Kcat$.)
\begin{enumerate}[(1)]
\item
The function $\vphi\typ (\DexpoKE)\fun (\DexpoKSE)$, $\vphi=\curry(\evalk)$
is bijective and monotonic.
(But need not be \cont.)\\
The translation $\vphi$ respects evaluation and curried functions:\\
$\eval(\vphi a,d)= \evalk(a,d)$, $\vphi\comp\curryk(f)=\curry(f)$.
\item
Let, furthermore, some $C\in \Kcat$ with some $c,c'\in \Cel$, $c\lex c'$ and $c\neq c'$,
and for all $f_1\lex f_2$ in $\DexpoKSE$
the function $h\typ C\pro D\fun E$ be in $\Kcat$ with 
$h(u,x)=f_1 x$ for $u\lex c$ and $h(u,x)=f_2 x$ else.
Let $g_1,g_2\typ \terminal \fun C$ be in $\Kcat$ with $g_1\bot = c$ and $g_2 \bot=c'$.\\
Then $\vphi^{-1}$ is monotonic, so that $\vphi$ is an order isomorphism.
(But need not be a $\Dlubpo$-isomorphism.)
\end{enumerate}
\end{prop}
\proof
\textbf{(1)}
The proof is a detailed version of the first part of the proof of part (iii), lemma 5 of \cite{Smyth}.

\begin{pspicture}(-2,-0.5)(7,3)
\firstn{DE}{\DexpoKE}
\nextn{DE}{xk}{\E}{0.7}{\prok}
\nextn{xk}{D1}{\E}{0.5}{D}
\nextn{DE}{T}{\N}{2}{\terminal}
\nextn{xk}{x}{\N}{2}{\pro}
\nextn{D1}{D2}{\N}{2}{D}
\nextn{D2}{E}{\E}{2}{E}
\ncline{->}{T}{DE}\Bput{$\curryk(f')$}
\ncline{->}{D2}{D1}\Aput{$\id$}
\ncline{->}{D2}{E}\Aput{$f'$}
\ncline{->}{D1}{E}\Bput{$\evalk$}
\end{pspicture}

$\vphi$ is \textbf{surjective}:\\
Let $f\in \DexpoKSE$.\\
Let $f'\typ\terminal\pro D\fun E$, with $f'y=f(\prot y)$, $f'=f\comp \prot$ is in $\Kcat$.\\
Let $h=(\curryk(f'))\bot \in (\DexpoKE)$. We calculate:
\begin{align*}
(\vphi h)x &= \evalk (h,x)\\
&= \evalk((\curryk(f'))\bot,x)\\
&= f'(\bot,x)\\
&= fx.
\end{align*}
So $\vphi h=f$.

$\vphi$ is \textbf{injective}:\\
Let $h'\in \DexpoKE$ with $\vphi h'=f$.
We have to show that $h'=h$ above.\\
Let $g\typ \terminal \fun (\DexpoKE)$ with $g\bot=h'$, $g$ is in $\Kcat$ by hypothesis.\\
We show that $\evalk\comp(g\pro \id)=f'$:
\begin{align*}
f'y &= f(\prot y)\\
&=(\vphi h')(\prot y)\\
&=\evalk(h',\prot y)\\
&=(\evalk\comp(g\pro \id))y.
\end{align*}
By the uniqueness of $\curryk(f')$ in $\Kcat$,
it must be $g=\curryk(f')$.\\
Therefore $h'=g\bot=\curryk(f')\bot =h$, as desired.\\
\textbf{(2)}
The proof is an adaptation of the second part of the proof of part (iii), lemma 5 of \cite{Smyth}.
\qed

We now come to situations, where a subcategory $\Kcat$ of $\Dlubpo$ has a categorical
(terminal, product, exponent) object $D$ and also the normal (terminal, product, general function)
dlubpo $D'$ with a certain morphism $\vphi\typ D\fun D'$,
and where this implies that $\vphi$ is an isomorphism.
For the terminal and product we can give two propositions where we generalize
from $\Dlubpo$ to an arbitrary category $\Ccat$.

\begin{prop}
Let $\Ccat$ be a category and $\Kcat$ a subcategory of $\Ccat$.
\begin{enumerate}[(1)]
\item
If $\Ccat$ has a terminal object $\terminal$, and $\Kcat$ a terminal object $\terminal_{\Kcat}$,
and $\Kcat$ contains $\terminal$ and a morphism $\vphi\typ \terminal_{\Kcat}\fun \terminal$,
then $\vphi$ and the unique (in $\Kcat$) $\psi\typ \terminal \fun \terminal_{\Kcat}$
are an isomorphism in $\Kcat$.
\item
Let $D,E\in \Kcat$.
If $\Ccat$ has a product $D\pro E$, with $\proi,\prot,\lpro\rpro$,
and $\Kcat$ a product $D\prok E$, with $\proik,\protk,\lpro\rpro_{\Kcat}$,
and $\Kcat$ contains $D\pro E$, $\proi,\prot$ and the morphism
$\vphi\typ D\prok E\fun D\pro E$, $\vphi=\lpro \proik,\protk\rpro$,\\
then $\vphi$ and $\psi\typ D\pro E\fun D\prok E$, $\psi=\lpro\proi,\prot\rpro_{\Kcat}$,
are an isomorphism in $\Kcat$.\\
The translation $\vphi$ respects projections and pairing functions.
\end{enumerate}
\end{prop}
\proof
The proofs are easy and like the known proofs of the isomorphism of terminal objects
\resp products in a single category.
\qed

The situation for exponents is special and more complicated:

\begin{prop}\label{p:expo2}
Let $\Kcat$ be a subcategory of $\Dlubpo$ with the normal terminal object $\terminal$
and all normal categorical products $\pro$. Let $D,E\in \Kcat$.\\
Let $\Kcat$ have a categorical exponent $\DexpoKE$ of $D,E$ with evaluation 
$\evalk$, curried morphisms $\curryk(f)$ 
and all constant functions $\la x.f\typ \terminal\fun(\DexpoKE)$, for $f\in \DexpoKEel$.\\
If $\Kcat$ also contains the general function dlubpo (relative to $\Kcat$) $\DexpoKSE$
with the morphism $\eval$,
then\\
$\vphi\typ (\DexpoKE)\fun(\DexpoKSE)$, $\vphi=\curry(\evalk)$ and\\
$\psi\typ (\DexpoKSE)\fun(\DexpoKE)$, $\psi=\curryk(\eval)$ are an isomorphism in $\Dlubpo$.\\
$\psi$ is, of course, in $\Kcat$.
If $\vphi$ is also in $\Kcat$, then the isomorphism is also in $\Kcat$.\\
The translation $\vphi$ respects evaluation and curried functions:\\
$\eval(\vphi a,d)= \evalk(a,d)$, $\vphi\comp\curryk(f)=\curry(f)$.
\end{prop}
\proof
By proposition \ref{p:expo1}(1), $\vphi$ is bijective and monotonic.
We have to prove that $\vphi$ is also continuous,
and that $\psi$ is the inverse of $\vphi$.

\begin{pspicture}(-2,-0.5)(7,3)
\firstn{DE}{\DexpoKSE}
\nextn{DE}{xk}{\E}{0.7}{\pro}
\nextn{xk}{D1}{\E}{0.5}{D}
\nextn{DE}{T}{\N}{1.9}{\DexpoKE}
\nextn{xk}{x}{\N}{2}{\pro}
\nextn{D1}{D2}{\N}{2}{D}
\nextn{D2}{E}{\E}{2}{E}
\ncarc[arcangle=20]{<-}{T}{DE}\Aput{$\psi$}
\ncarc[arcangle=20]{<-}{DE}{T}\Aput{$\vphi$}
\ncline{<->}{D2}{D1}\Aput{$\id$}
\ncline{->}{D2}{E}\Aput{$\evalk$}
\ncline{->}{D1}{E}\Bput{$\eval$}
\end{pspicture}

For all $f\in \DexpoKSE$, $d\in\Del$,
it is 
\begin{align*}
\evalk((\psi f),d) &= \evalk(\curryk(\eval)f,d)\\
&= \eval(f,d).\\
\text{Then } \vphi(\psi f) &= \la d\in D. \evalk(\psi f,d)\\
&= \la d\in D. \eval(f,d) = f.
\end{align*}
Therefore $\psi$ is the inverse of $\vphi$, as $\vphi$ is bijective.

To show that $\vphi$ is \cont, let $A\nlub a$ in $\DexpoKE$.
We have to show that $\fset{\vphi} A\nlub \vphi a$.\\
So let $B\sub (\DexpoKSE)\pro\Del$ directed with $\fset{\proi}B=\fset{\vphi} A$
and $\fset{\prot}B\nlub d$.\\
We have to show that $\fset{\eval} B\nlub \eval(\vphi a,d)$.\\
Let $C=\set{(\psi f,b)}{(f,b)\in B}$. $C$ is directed, as $B$ is directed and $\psi$ is monotonic.\\
It is $\fset{\proi}C=A$ and $\fset{\prot}C=\fset{\prot} B\nlub d$. We calculate:
\begin{align*}
\fset{\eval} B &= \set{\evalk(\psi f,b)}{(f,b)\in B}\\
&= \fset{\evalk} C\\
&\nlub \evalk(a,d)\\
&= \evalk(\psi(\vphi a),d)\\
&= \eval(\vphi a,d).
\end{align*}
So $\vphi$ is an isomorphism in $\Dlubpo$.
\qed

\begin{thm}\label{t:dlubpo}
In $\Dlubpo$ $\terminal$ is the terminal object and $D\pro E$ is the categorical product with
$\proi,\prot$ and $\lpro f,g\rpro$.\\
If there is an exponent of $D,E$ in $\Dlubpo$,
then it is $\DexpoSE$ with evaluation $\eval$ and curried functions $\curry(f)$.\\
$\Dlubpo$ is not a ccc: there are dlubpos $D,E$ such that $\DexpoSE$ is not the exponent,
\ie there is some $f\typ C\pro D\fun E$ such that $\curry(f)$ is not \cont.
\end{thm}
\proof
The first sentence follows from the definitions of $\terminal$ and $D\pro E$.
The second follows from proposition \ref{p:expo2}.\\
Here is the counter-example in figure \ref{f:DEC}.

\psset{unit=5mm}
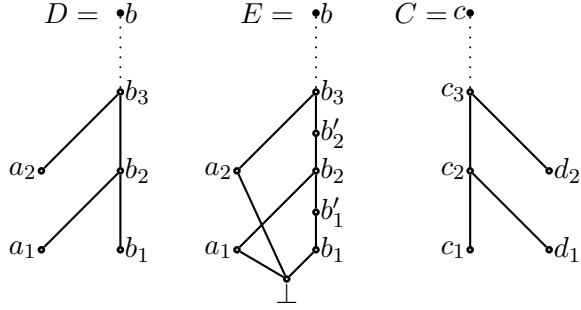
\begin{figure}[htb]
\begin{pspicture}(-2.7,-1.3)(13,7.0)
\firsto{b1}{\E}{b_1}
\slineo{b1}{b2}{\N}{2}{\E}{b_2}
\slineo{b2}{b3}{\N}{2}{\E}{b_3}
\dlineon{b3}{b}{\N}{2}{\E}{b}
\nextn{b}{D}{\W}{1.2}{D=}
\nexto{b1}{a1}{\W}{2}{\W}{a_1}
\nexto{b2}{a2}{\W}{2}{\W}{a_2}
\sline{a1}{b2}
\sline{a2}{b3}
\nexto{b1}{a1n}{\E}{3}{\W}{a_1}
\nexto{a1n}{b1n}{\E}{2}{\E}{b_1}
\slineo{b1n}{b1nn}{\N}{0.9}{\E}{b_1'}
\slineo{b1nn}{b2n}{\N}{1}{\E}{b_2}
\slineo{b2n}{b2nn}{\N}{0.9}{\E}{b_2'}
\slineo{b2nn}{b3n}{\N}{1}{\E}{b_3}
\dlineon{b3n}{bn}{\N}{2}{\E}{b}
\nextn{bn}{E}{\W}{1.2}{E=}
\nexto{b2n}{a2n}{\W}{2}{\W}{a_2}
\sline{a1n}{b2n}
\sline{a2n}{b3n}
\nexto{b1n}{bot}{\SW}{1}{\S}{\bot}
\sline{bot}{a1n}
\sline{bot}{b1n}
\sline{bot}{a2n}
\nexto{b1n}{c1}{\E}{4}{\W}{c_1}
\slineo{c1}{c2}{\N}{2}{\W}{c_2}
\slineo{c2}{c3}{\N}{2}{\W}{c_3}
\dlineon{c3}{c}{\N}{2}{\W}{c}
\nextn{c}{C}{\W}{1.2}{C=}
\nexto{c1}{d1}{\E}{2}{\E}{d_1}
\nexto{c2}{d2}{\E}{2}{\E}{d_2}
\sline{d1}{c2}
\sline{d2}{c3}
\end{pspicture}
\caption{Example theorem \ref{t:dlubpo}, dlubpos $D,E,C$}\label{f:DEC}
\end{figure}
\psset{unit=1cm}

In the dlubpos $D,E,C$ the order is the reflexive, transitive closure of the relations
specified below.
They also contain all trivial natural lubs,
\ie all natural lubs $A\nlub a$ where $a\in A$.
So also the subcategory of the dlubpos with all trivial natural lubs is not a ccc.

$D$ is the dlubpo with\\
$\Del=\aiset\join\biset\join \{b\}$,\\
for all $i$: $a_i\lex b_{i+1}$, $b_i\lex b_{i+1}$,
$b$ greatest element,\\
$D'=\aiset\join\biset$,
$D'\nlub b$.

$E$ is the dlubpo with:\\
$\Eel=\aiset\join\biset\join \bipset\join\{b\}$,\\
for all $i$: $a_i\lex b_{i+1}$, $b_i\lex b_i'\lex b_{i+1}$,
$b$ greatest element, $\bot$ least element,\\
for all directed $S\sub\Eel$ with the lub $b$ and some $b_i'\in S$ it is $S\nlub b$.

$C$ is the dlubpo with:\\
$\Cel=\diset\join\ciset\join\{c\}$, it is isomorphic to $D$,\\
for all $i$: $d_i\lex c_{i+1}$, $c_i\lex c_{i+1}$,
$c$ greatest element,\\
$C'=\diset\join\ciset$,
$C'\nlub c$.

We define a \cont\ function $f\typ C\pro D\fun E$ by defining $\curry(f)$.
In the following, we abbreviate $\curry(f)x=\overline{x}$ for all $x\in\Cel$.
For all $n\geq 1$ we define:
\begin{align*}
\cnb a_1 &= \bot & \dnb a_1 &= \bot\\
\cnb a_i &= a_{i-1} \text{ for } 2\leq i\leq n & \dnb a_i &= a_{i-1} \text{ for } 2\leq i<n\\
\cnb a_i &= \bot \text{ for } i>n & \dnb a_i &= \bot \text{ for } i\geq n \\
\cnb b_i &= b_i' \text{ for } i<n & \dnb b_i &= b_i' \text{ for } i<n\\
\cnb b_i &= b_{n-1}' \text{ for } i\geq n & \dnb b_i &= b_n \text{ for } i\geq n\\
\cnb b &= b_{n-1}' & \dnb b &= b_n\\[3mm]\db
\cb a_1 &= \bot\\
\cb a_i &= a_{i-1} \text{ for } i\geq 2\\
\cb b_i &= b_i' \text{ for } i\geq 1\\
\cb b &= b
\end{align*}

\begin{figure}[htb]
\begin{pspicture}(-0.7,-0.2)(2.5,3.5)
\firsto{c1}{\W}{\overline{c_1}}
\slineo{c1}{c2}{\N}{1}{\W}{\overline{c_2}}
\slineo{c2}{c3}{\N}{1}{\W}{\overline{c_3}}
\dlineon{c3}{c}{\N}{1}{\W}{\cb}
\nexto{c1}{d1}{\E}{1}{\E}{\overline{d_1}}
\slineo{d1}{d2}{\N}{1}{\E}{\overline{d_2}}
\slineo{d2}{d3}{\N}{1}{\E}{\overline{d_3}}
\sline{c1}{d2}
\sline{c2}{d3}
\sline{d1}{c2}
\sline{d2}{c3}
\end{pspicture}
\caption{Example theorem \ref{t:dlubpo}, functions $\bar{x}=\curry(f)x$}\label{f:barx}
\end{figure}
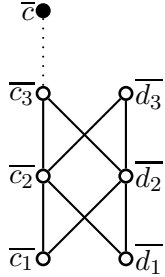

For all $x\in\Cel$, the function $\overline{x}$ is monotonic.
Furthermore, for all $n$ it is
$\cnb\lex \overline{c_{n+1}}$,
$\dnb\lex \overline{c_{n+1}}$,
$\cnb\lex \cb$,
also $\dnb\lex \overline{d_{n+1}}$
and  $\cnb\lex \overline{d_{n+1}}$,
so that we get the diagram figure \ref{f:barx}.

From all this follows that $f$ is monotonic.\\
$f$ is also \cont:
let $A\sub \Cel\pro\Del$ directed with $\fset{\proi} A\nlub e_1$ and $\fset{\prot} A\nlub e_2$.
We have to show that $\fset{f}A\nlub f(e_1,e_2)$.
The proof is by going through the cases, 
we leave out the cases where one (or two) of the natural lubs is a trivial natural lub.\\
The interesting case is $\fset{\proi} A=C'\nlub c$ and $\fset{\prot} A=D'\nlub b$:\\
For every $n$ there is some $(c_n,y)\in A$ and some $(x,b_n)\in A$.\\
As $A$ is directed, there must be some $(c_i,b_j)\in A$ with $i\geq n$ and $j\geq n$,
then it is $b_{n-1}'\lex f(c_i,b_j)$.
Therefore $b$ is the lub of $\fset{f} A$.
Also $f(c_i,b_j)=b_k'$ for some $k$.
Therefore $\fset{f}A\nlub b=f(c,b)$.
So $f$ is \cont.

We now prove that $\curry(f)$ is not \cont:\\
It is $C'\nlub c$ in $C$.
We define $\overline{C'}=\fset{(\curry(f))}C'$.
It is $\cb=\curry(f)c$.
We prove that not $\overline{C'}\nlub \cb$ in $\DexpoSE$.\\
Let $A\sub \DexpoSEel \pro \Del$ be the set $A=\{(\cnb,a_n)\}_{n\geq1} \join \{(\dnb,b_n)\}_{n\geq 1}$.\\
$A$ is directed, as for all $n$ it is
$(\cnb,a_n)\lex (\overline{d_{n+1}},b_{n+1})$
and $(\dnb,b_n)\lex (\overline{d_{n+1}},b_{n+1})$.\\
It is $\fset{\proi}A=\overline{C'}$ and $\fset{\prot}A=D'\nlub b$,
but $\fset{\eval}A = \{\bot\}\join \aiset\join\biset$ does not have
$\eval(\cb,b)=b$ as natural lub,
as $\fset{\eval}A$ does not contain any $b_i'$.
\qed

We now come to the problem of finding a large sub-ccc of $\Dlubpo$
(that uses the general exponents $\DexpoSE$).
We conjecture that it is possible to find such a category that is the largest among those
sub-cccs that are generated by some kind of invariant lub-rule class,
see the following section \ref{s:lubrule},
analogous to the result of theorem \ref{t:S10}  for subcategories that use the
(pointwise) exponent $\DexpoE$.
But it seems that this comes at a price: 
the needed axiom will be rather complicated,
it encodes the condition that curried functions are \cont.
And the definition of the notion of lub-rule class has to be augmented.
As we think that such a result will not be worth the effort,
we do not follow this,
but show here that the simple axiom S5 is sufficient for cartesian closedness,
though not in any sense maximal.

\begin{defi}
A dlubpo $D$ is an \defin{S5-dlubpo} if it fulfills axiom S5 for directed sets:
If $X\sub Y\sub\Del$, $X,Y$ directed, $Y\nlub y$ and $Y\lset X$, then $X\nlub y$.
\end{defi}

Note that from $X\sub Y\sub\Del$, $Y$ directed and $Y\lset X$ 
follows that $X$ is also directed.
Therefore by axiom S5 we can deduce from natural lubs of directed sets only 
natural lubs of sets that are directed again.
So we get the same class of dlubpos when we demand axiom S5 for general sets in the definition,
contrary to the situation for natural domains.

\begin{prop}\label{p:S5}
Let $D,E,C$ be dlubpos.
\begin{enumerate}[(1)]
\item
The terminal dlubpo $\terminal$ fulfills S5.
\item
If $D,E$ fulfill S5, then $D\pro E$ fulfills S5.
\item 
If $E$ fulfills S5, then $\DexpoSE$ fulfills S5.
\item
If $E$ fulfills S5, then for all $f\typ C\pro D\fun E$ it is $\curry(f)\typ C\fun (\DexpoSE)$ \cont.
\end{enumerate}
\end{prop}
\proof
\textbf{(1)} is clear.\\
\textbf{(2)}
Let $X\sub Y \sub |D\pro E|$ directed, $Y\nlub y$ and $Y\lset X$.
Then $\fset{\proi}X\sub \fset{\proi}Y$ directed,
$\fset{\proi}Y\nlub \proi y$
and $\fset{\proi}Y\lset \fset{\proi} X$,
so $\fset{\proi}X\nlub \proi y$.
Also $\fset{\prot}X\nlub \prot y$. Therefore $X\nlub y$.\\
\textbf{(3)}
Let $F\sub G\sub \DexpoSEel$ directed, $G\nlub g$ and $G\lset F$.
We have to show $F\nlub g$.\\
Let $A\sub \DexpoSEel\pro \Del$ directed with $\fset{\proi}A=F$ and $\fset{\prot}A\nlub d$.
We have to show $\fset{\eval}A\nlub gd$.\\
Let $B=G\pro (\fset{\prot}A)$.
$B$ is directed, so $\fset{\eval}B\nlub gd$.\\
For every $(h,a)\in B$ there is some $(h',y)\in A$ with $h\lex h'$,
and some $(x,a)\in A$.\\
As $A$ is directed, there is some $(h'',a')\in A$ with $(h,a)\lex (h'',a')$.\\
So we get $B\lset A$, therefore $\fset{\eval}B\lset \fset{\eval}A$.\\
From $A\sub B$ follows $\fset{\eval}A\sub \fset{\eval}B$.\\
From all this follows $\fset{\eval}A\nlub gd$, as $E$ fulfills S5.\\
\textbf{(4)}
Let $C'\nlub c$ in $C$. Let $\overline{C'}=\fset{(\curry(f))}C'$ and $\cb=\curry(f)c$.\\
We have to show $\overline{C'}\nlub \cb$ in $\DexpoSE$.\\
Let $A\sub \DexpoSEel\pro \Del$ directed with $\fset{\proi}A=\overline{C'}$
and $\fset{\prot}A\nlub d$.\\
We have to show $\fset{\eval}A\nlub \cb d$.\\
Let $B\sub\Cel\pro\Del$ with $B=C'\pro(\fset{\prot}A)$.
$B$ is directed and $B\nlub(c,d)$, so $\fset{f}B\nlub f(c,d)$.\\
For every $(b,a)\in B$ there is some $(\curry(f)b,y)\in A$ and some $(x,a)\in A$.\\
As $A$ is directed, there is some $(g,a')\in A$ with
$(\curry(f)b,y)\lex (g,a')$ and $(x,a)\lex (g,a')$,
so $f(b,a)\lex \eval(g,a')$.\\
Therefore we get $\fset{f}B\lset \fset{\eval}A$.\\
Together with $\fset{\eval}A\sub \fset{f}B$ and $\fset{f}B\nlub \cb d$
we get $\fset{\eval}A\nlub \cb d$,
as $E$ fulfills S5.
\qed

It follows this theorem:

\begin{thm}
The full subcategory $\mathbf{S5dlubpo}$ of $\Dlubpo$, of all S5-dlubpos,
is cartesian closed,
with the normal terminal object $\terminal$,
the normal products $D\pro E$,
the general exponents $\DexpoSE$,
and the corresponding morphisms.
\qed
\end{thm}

%% file: T5.tex
\section{Lub-rule classes and closed dlubpos}\label{s:lubrule}

In this section we set up the general frame of lub-rule systems and classes,
in which we can describe  axiom systems that are like those of
Sazonov natural domains and dlubpos.
In this frame we can give a precise definition of the completeness of the axiom system;
and in section \ref{s:incomplete} it will turn out that the axioms of
Sazonov natural domains are not complete.

Let us have a look at an example axiom from section \ref{s:sazonov}:\\
(S6 cofinality) If $X,Y\sub \Del$, $X\nlub x$ and $X\lset Y\lset x$, then $Y\nlub x$.\\
Here we first have as hypothesis of the axiom a subset with a natural lub, $X\nlub x$,
and a further subset $Y$ with an order-theoretic configuration, $X\lset Y\lset x$.
From this configuration follows $\Lubex Y=x$,
and $Y\nlub x$ is the conclusion of the axiom.

The important property of the configuration, $X\lset Y\lset x$,
is that it is invariant against monotonic functions $f$ into another 
partial order $E$: it is $\fsetf X\lset \fsetf Y\lset fx$.
If $f$ also respects the lub $\Lubex X=x$, \ie $\Lubex \fsetf X=fx$,
then also in $E$ we can draw the conclusion $\Lubex \fsetf Y=fx$.
We can say this property of invariance means that the conclusion $\Lubex Y=x$,
and $Y\nlub x$ in the axiom, is ``necessary''.

All the axioms we have seen so far are of this form.
(Except axiom S1 which we do not count as proper axiom.)
They are necessary deductions of a lub from some set of natural lubs in an
invariant order-theoretic configuration.
We call these axioms ``valid''.

Here is an axiom that is not valid:
If $X\sub \Del$ and $\Lubex X=x$, then $X\nlub x$.
It is not valid, because there are monotonic functions $f$ from $D$ that do not
respect $\Lubex X=x$.
The monotonic function $f$ needs to respect only the natural lubs of the 
hypothesis of the axiom, and here there are none.


We now formalize these intuitions.
Our results should also be applicable to domains that use natural lubs of \emph{general}
subsets, like natural domains.
For this we define lub partial orders as general structures,
only for use in this section.
(Further below we will go over from lubpos to dlubpos.)

\begin{defi}
A structure $D=(\Del,\lexD,\nlubD)$ is a \defin{lub partial order (lubpo)} if\\
$\Del$ is a set of \defin{elements},\\
$\lexD$ is a partial order on $\Del$,\\
$\nlubD\;\sub \Potlub(D)\pro\Del$ is the \defin{(natural) convergence relation}
with $\Lubex A=a$ for $A\nlub a$.
(Remember that $\Potlub(D)$ is the set of subsets of $\Del$ that have a lub in $\lexD$.)\\
If $(\Del,\lexD)$ is a partial order and $P\sub\Potlub(D)$,
then $\nlubP\;\sub \Potlub(D)\pro\Del$ is the relation with
$A\nlubP(\Lubex A)$ for all $A\in P$,
so that $(\Del,\lexD,\nlubP)$ is the lubpo corresponding to the set $P$ of natural subsets.
\end{defi}

A lub-rule models an instance of the application of an axiom:

\begin{defi}
A \defin{lub-rule} on a partial order $D$ is a triple $(D,P\ruleto A)$
with $P\sub\Potlub(D)$ and $A\in \Potlub(D)$.
$P$ is called the \defin{pattern} of the lub-rule and $A$ its \defin{result}.

A lub-rule $(D,P\ruleto A)$ is \defin{valid} if for every partial order $E$
and every monotonic function $f\typ D\fun E$
that respects the lubs of the elements of $P$ 
(\ie $\all{B\in P}f(\Lubex B)=\Lubex(\fset{f}B)$),
$f$ respects also the lub of $A$
(\ie $f(\Lubex A)=\Lubex(\fset{f} A)$).

A lubpo $E$ \defin{fulfills} a lub-rule $(D,P\ruleto A)$ if:\\
If $(\Eel,\lexE)=D$ and all elements of $P$ are natural in $E$,
then $A$ is natural in $E$.

A \defin{lub-rule system} $\Rcal$ on a partial order $D$ is a set of pairs (rules)
$(P\ruleto A)$ such that $(D,P\ruleto A)$ is a valid lub-rule on $D$.
This is a rule system in the sense of definition \ref{d:rulesystem},
so we get derived lub-rules from the lub-rule system.
(Proposition \ref{p:derived} below shows that the derived rules are valid.)
In this context we implicitely understand every rule $(P\ruleto A)$ of $\Rcal$
as the lub-rule $(D,P\ruleto A)$.

\noindent This lub-rule system $\Rcal$ is \defin{complete} if every valid lub-rule on $D$
is derivable from it.

\noindent A \defin{lub-rule class} is a class of valid lub-rules.
(They can be on different partial orders.)

\noindent A lub-rule class $\Rcal$ \defin{generates} the class of lubpos that fulfill all lub-rules of $\Rcal$.

A lub-rule class $\Rcal$ \defin{induces} on every partial order $D$ the lub-rule system of
all $(P\ruleto A)$ with $(D,P\ruleto A)\in \Rcal$.
$\Rcal$ is \defin{complete} if all these induced lub-rule systems are complete.
A lub-rule is \defin{derived} from $\Rcal$,
if it is derived in one of the induced lub-rule systems.

A lub-rule class $\Rcal$ is \defin{invariant} if for every lub-rule $(D,P\ruleto A)\in \Rcal$,
every partial order $E$ and every monotonic function $f\typ D\fun E$
that respects the lubs of the elements of $P$,
also the lub-rule $(E,\fset{(\fsetf)}P\ruleto \fset{f}A)$
is derivable in the lub-rule system induced by $\Rcal$ on $E$.
(It is $\fset{(\fsetf)} P = \set{\fsetf B}{B\in P}$.)
(The last is a valid lub-rule as the first is it.)
\end{defi}

\begin{prop}
If we have a lubpo $E$ and a lub-rule system $\Rcal$ on $(\Eel,\lexE)$
such that $E$ fulfills every rule of $\Rcal$,
then $E$ fulfills also every derived rule of $\Rcal$.\\
Every complete lub-rule class is invariant.
\qed
\end{prop}

Let us demonstrate the translation from the axiom system to the lub-rule class for
the axiom S6 (cofinality).
It is translated to the lub-rule class of all lub-rules $(D,\{X\}\ruleto Y)$
such that $D$ is a partial order, $X\sub\Del$ with $\Lubex X=x$ for some $x$,
and $X\lset Y\lset x$.
All these are valid lub-rules.
The other axioms S3 and S7 of natural domains are likewise translated,
so that we get an invariant lub-rule class $\Scal$
for the axioms of Sazonov natural domains.
(The proof is immediate and boring.)

\subsection*{Philosophical significance: the quest for intension by partial extensionalization}

Let us reflect on what we have done so far.
The following is not mathematics, it is the thoughts that I had when discovering
this piece of mathematics.

Since the Logic of Port-Royal (1662),
philosophical logic makes the distinction between two kinds of meaning of a concept:
the \emph{intension} (or content),
\ie the predicates the concept uses to describe the things which fall under it,
and the \emph{extension},
\ie the class of all these things.

Mathematics, as based on set theory, is always extensional,
\ie we use \emph{intensions} (like expressions, axioms),
to describe, define \emph{extensions} (like sets, classes).
We always talk in intensions about extensions.


But what can we do when we want to talk \emph{in mathematics} about an intension?
We have to \emph{extensionalize} an aspect of the intension,
we have to introduce new extensions.
This may occur at different levels, the intension covers a whole spectrum of them.
We are always sure to have the lowest level,
\ie the purely syntactic text on which the intension is founded,
and the highest level, 
the extension that the intension describes.
We call the last one the \emph{full extensionalization},
and a level properly between the lowest and the highest a \emph{partial extensionalization}.

In our concrete case we have as intension an axiom system for lubpos,
which defines as extension a class of lubpos.
We deliberately leave open the formal logic of the axiom system,
so that we describe the following translation informally.
The axiom system must be amenable to this translation to a lub-rule class,
which is our partial extensionalization.
A lub-rule abstracts a certain form of inference from the axiom, in extensional form.
It must be defined that the partial extensionalization describes the full one,
which is done by our definition of a lub-rule class generating a class of lubpos.
We have gained new objects,
the lub-rule classes, which can be used to analyse classes of lubpos.

There might be similar examples in the literature, but I know only the following one.
Sometimes the partial extensionalization works by forming equivalence classes
of the syntactic expressions of the intensions.
This is the case for the formalization of the notion of algorithm given by Noson
Yanofsky \cite{Yanofsky}.
Here the intensional objects are (primitive recursive) programs,
and the full extensions are the functions they describe.
The algorithms are defined as certain equivalence classes of programs.

\bigskip

Now we introduce the important cl-rule system on a lubpo and its closure operation,
and the lub-completion of a lubpo.
These procedures appear in \cite{Courcelle/Raoult}
and are used there for various completions of partial orders that respect lubs
that are already fixed.
In this section  we use them to give a characterization of valid lub-rules;
this will be used in section \ref{s:incomplete} to prove the
incompleteness of the Sazonov axioms $\Scal$.

\begin{defi}[Courcelle/Raoult, proof of theorem 1 in \cite{Courcelle/Raoult}]\label{d:cl}
Let $D$ be a lubpo.
The \defin{cl-rule system} on $D$ is the rule system
(in the sense of def.\ \ref{d:rulesystem}) $\ruleto$ on $\Del$ given by the rules:\\
for all $a,b\in \Del$ with $b\lex a$: $\{a\}\ruleto b$,\\
for all $A\sub \Del$, $a\in \Del$ with $A\nlub a$: $A\ruleto a$.\\
The closure operator of the cl-rule system is called $\clD$.\\
An $A\sub\Del$ is \defin{closed} if it is closed under $\clD$.\\
(We remark that closed sets are a generalization of subsets of dcpos
that are closed \wrt the Scott topology.)\\
If we have a lubpo given this way: $D=(\Del,\lexD,\nlubP)$,
with its set of natural sets $P$,
then we abuse notation and write $\clP$ for $\clD$,
if $\Del$ and $\lexD$ are clear from the context.\\
$\Dcl$ is the set of all closed subsets of $D$.\\
The \defin{lub-completion} of $D$ is $\lubcomp(D)=(\Dcl,\lex)$,
with $\lex$ the inclusion $\sub$.\\
It is a complete lattice with $\Lubex \bar{A}= \clD(\Join \bar{A})$, for $\bar{A}\sub \Dcl$.\\
There is an embedding $\innD\typ \Del\fun\Dcl$ defined by
$\innD d = \clD\{d\}=\set{x\in D}{x\lex d}$.
\end{defi}

\begin{prop}[Courcelle/Raoult, proof of theorem 1 in \cite{Courcelle/Raoult}]\label{p:in}
The embedding $\innD$ is an order embedding,
\ie it is monotonic, injective, and the reverse is monotonic. It maps natural lubs in $D$ to lubs in $\Dcl$,
\ie for $A\nlub a$ in $D$ it is $\innD a = \Lubex(\fset{\innD} A)= \clD A$.
\end{prop}
\proof
$\innD$ and its reverse are clearly monotonic.\\
Now let $A\nlub a$ in $D$.
It is $A\sub \Join(\fset{\innD} A)$,
then $a\in \cl(\Join(\fset{\innD} A)) = \Lubex(\fset{\innD} A)$,\\
so $\innD a \sub \Lubex(\fset{\innD} A)$.\\
For the reverse, it is $\innD b\sub\innD a$ for all $b\in A$.
\qed

\begin{prop}\label{p:valid}
Let $(D,P\ruleto A)$ be a lub-rule and 
$E=(\Del,\lexD,\nlubP)$ be the corresponding lubpo
with the set $P$ of natural subsets.
Let $\Lubex A=a$.\\
$(D,P\ruleto A)$ is valid    $\ifff$  $a\in \clE(A)$, also written $a\in \clP(A)$.
\end{prop}
\proof
$\ifthen$:
We have the embedding $\inn\typ (\Eel,\lex)\fun \lubcomp(E)=(\hat{E},\lex)$.\\
By proposition \ref{p:in}, the function $\inn$ respects the natural lubs of the elements of $P$.
Whence by validity it also respects the lub of $A$, so $\inn(\Lubex A)=\Lubex(\fset{\inn} A)$.\\
Further we get $\Lubex(\fset{\inn} A) = \clE(\Join(\fset{\inn} A))=\clE(A)$,
thus $a\in\clE(A)$.

\noindent $\thenif$:
Let $F=(|F|,\lex_F)$ be a partial order and
$f\typ D\fun F$ be a monotonic function that respects the lubs of the elements of $P$.
We have to show: $fa=\Lubex \fset{f}A$.\\
We prove:
For all $x\in \clE(A)$,
for all upper bounds $b$ of $\fset{f}A$, it is $fx\lex b$,
by induction on the deduction of $x\in \clE(A)$ in the cl-rule system on $E$.\\
(1) This is true for $x\in A$.\\
(2) Let $x$ be deduced from $y\in \clE(A)$ by $x\lex y$.
Then $fx\lex fy\lex b$.\\
(3) Let $x$ be deduced from $Y\sub\clE(A)$ by $Y\nlub x$.
Then $fx=\Lubex(\fsetf Y)\lex b$.\\
It is clear that $fa$ is an upper bound of $\fsetf A$.
The proved property for $x=a$ shows that it is the least upper bound.
\qed

\begin{prop}\label{p:derived}
Let $\Rcal$ be a lub-rule system on a partial order $D$.
Every derived rule $P\ruleto A$ of $\Rcal$ is valid.
\end{prop}
\proof
Let D1 be a deduction of $A$ from $P$ in the lub-rule system $\Rcal$.\\
We prove by induction on D1 that for every node of D1 labelled with some deduced $A'$,
with $\Lubex A'=a'$,
it is $a'\in \clP(A')$.\\
(1) This is clear for $A'\in P$.\\
(2) Let $A'$ be deduced from some $P'$ at this node, $P'\ruleto A'$.\\
From proposition \ref{p:valid} follows that $a'\in \cl_{P'}(A')$.\\
Let D2 be a deduction of $a'$ from $A'$ in the cl-rule system on $(\Del,\lexD,\nlub^{P'})$.\\
We prove by induction on D2 that for every node of D2 labelled with some deduced $x$,
it is $x\in\clP(A')$.\\
(2.1) This is clear for $x\in A'$.\\
(2.2) Let $x$ be deduced from $y$ by $x\lex y$.
It is $y\in\clP(A')$ by induction hypothesis,
so $x\in \clP(A')$.\\
(2.3) Let $x$ be deduced from $Y\in P'$ by $Y\nlub x$.\\
The elements of $P'$ were deduced in the deduction D1,
so by induction hypothesis of the outer induction on D1 it is $x\in \clP(A')$.

From the inner induction on D2 follows that $a'\in \clP(A')$.\\
From the outer induction on D1 follows that $\Lubex A\in\clP(A)$,
so $P\ruleto A$ is valid by proposition \ref{p:valid}.
\qed

The easiest way to get a complete lub-rule class is to take just all valid lub-rules,
which can be characterized by proposition \ref{p:valid}:
\begin{defi}
The \defin{canonical complete lub-rule class} $\Ccal$ is the class of all valid lub-rules.
This is the set of all lub-rules $(D,P\ruleto A)$ with $\Lubex A\in \clP(A)$.
\end{defi}

\bigskip

We are mainly interested in the natural sets that are directed, as continuity is based on them,
and regard the undirected natural sets only as intermediary (may be necessary)
steps in the deduction of naturality of directed sets,
see the discussion in section \ref{s:sazonov}.
So we go over from lubpos to dlubpos.

\begin{defi}
A lubpo that fulfills the singleton axiom S3 is called a \defin{lubpo with singletons}.
For a lubpo with singletons we have the \defin{dlubpo for} $D$, $\delta(D)= (\Del,\lexD,\nlub)$
with $\nlub$ the restriction of $\nlubD$ to directed subsets.

A lub-rule class $\Rcal$ is \defin{with singletons}
if $\Rcal$ has for all partial orders $D$ and all $x\in \Del$ the (valid) 
lub-rule $(D,\emptyset\ruleto \{x\})$.

A lub-rule $(D,P\ruleto A)$ is \defin{directed} if $A$ and the elements of $P$ are all directed.\\
For a lub-rule class $\Rcal$ with singletons,
$\delta(\Rcal)$ is the lub-rule class of all derived lub-rules of $\Rcal$ that are directed.\\
A lub-rule class $\Rcal$ with singletons \defin{generates} the class of all 
dlubpos $\delta(D)$ for $D$ a lubpo that fulfills the lub-rules of $\Rcal$.
$\Rcat$ is the full subcategory of this class in $\Dlubpo$.
\end{defi}

\begin{prop}
A lub-rule class $\Rcal$ with singletons generates the class of dlubpos
that fulfill all lub-rules of $\delta(\Rcal)$.
\end{prop}
\proof
We have to prove that very dlubpo $D$ that fulfills all lub-rules of $\delta(\Rcal)$
can be extended to a lubpo $D'=(\Del,\lexD,\nlub')$ such that 
$D'$ fulfills the lub-rules of $\Rcal$ and $\delta(D')=D$.\\
Take the set $P$ of all (directed) natural sets of $D$ and build the closure $P'$ of $P$
under the lub-rule system induced by $\Rcal$ on $(\Del,\lexD)$.
Take $\nlub'\,=\,\nlub^{P'}$.
\qed

If we use all valid lub-rules,
the deduction of (directed) natural lubs from existing ones is always in one step.
Thus we can define the class of dlubpos that is generated by $\Ccal$ by an axiom
on directed lubs that tests the condition of proposition \ref{p:valid}:
\begin{defi}
A dlubpo $D=(\Del,\lexD,\nlubD)$ is a \defin{closed dlubpo (cdlubpo)}
if it fulfills the axiom:\\
\textbf{(S9 closure)} If $A\sub\Del$ is directed with lub $a$
and $a\in \clD(A)$, then $A\nlubD a$.
\end{defi}
Note that this axiom, as it stands, does not describe a complete lub-rule class.
But its extension is exactly the class of dlubpos generated by the complete lub-rule class $\Ccal$.

\begin{prop}
A dlubpo $D$ is a cdlubpo $\ifff$ $D$ fulfills all valid directed lub-rules\\
$((\Del,\lexD),P\ruleto A)$.\\
The category $\Cdlubpo$ of cdlubpos is the category $\Ccal$-$\mathbf{cat}$.
\end{prop}
\proof
$\ifthen$:
Let $((\Del,\lexD),P\ruleto A)$ be a valid lub-rule of directed sets such that all elements
of $P$ are natural in $D$, and $\Lubex A=a$.
Then $a\in \clP(A)$ by proposition \ref{p:valid},
therefore also $a\in \clD(A)$.
Thus $A\nlub a$.

$\thenif$:
Let $A\sub \Del$ be directed with lub $a$ and $a\in \clD(A)$.
Let $P$ be the set of all natural (directed) subsets of $\Del$.
Then $((\Del,\lexD),P\ruleto A)$ is valid by proposition \ref{p:valid}. Thus $A\nlub a$.  
\qed

As a preparation for the next sections, we now give some properties of the classes of dlubpos
that are generated by (invariant) lub-rule classes.

\begin{defi}
Let $D$ be a dlubpo with the set $P$ of natural directed subsets,
and $\Rcal$ be a lub-rule class with singletons.
The $\Rcal$-completion $\Rcomp(D)$ of $D$ is the dlubpo $(\Del,\lexD,\nlub^{P'})$
with $P'$ the directed sets of the closure of $P$ under the lub-rule system 
induced by $\Rcal$.
\end{defi}

\begin{prop}\label{p:frcomp}
Let $D,E$ be dlubpos and $\Rcal$ be a lub-rule class with singletons.
\begin{enumerate}[(1)]
\item
$D$ is in the class of dlubpos generated by $\Rcal$ iff $\Rcomp(D)=D$.
\item
If $\Rcal$ is invariant, then every \cont\ function $f\typ D\fun E$ is also
a \cont\ function $f\typ \Rcomp(D)\fun \Rcomp(E)$.
\end{enumerate}
\end{prop}
\proof
(1) is clear.\\
(2) Let $D'=\Rcomp(D)$ and $E'=\Rcomp(E)$.\\
We have to prove: For every $A\nlub a$ in $D'$ it is
$\fsetf A\nlub fa$ in $E'$.\\
There is a deduction of $A$ from the set $P$ of natural directed sets of $D$ 
in the lub-rule system induced by $\Rcal$.
We prove by induction on this deduction,
that in every node for the deduced $A'$ (with lub $a'$)
it is $\Lubex \fsetf A'= fa'$ and $\fsetf A'$ is in the closure of
$\fsset{f}P$ in the lub-rule system of $\Rcal$ on $E$.\\
This is clear for $A'\in P$, as $f\typ D\fun E$ is \cont.\\
Now let $A'$ be deduced from the set of subsets $P'$
by the (valid) lub-rule $(D,P'\ruleto A')$.\\
By induction hypothesis $f$ respects the lubs of the elements of $P'$,
so also the lub of $A'$: $\Lubex \fsetf A' = fa'$.\\
By induction hypothesis for all $B\in P'$, $\fsetf B$ is in the closure of $\fsset{f} P$.
As $\Rcal$ is invariant, $(E,\fsset{f}P'\ruleto \fsetf A')$ is a derived lub-rule
in the lub-rule system of $\Rcal$.
Therefore also $\fsetf A'$ is in the closure of $\fsset{f}P$.

From the proved hypothesis follows that $\fsetf A \nlub fa$.
\qed

\begin{thm}
Let $\Rcal$ be an invariant lub-rule class with singletons.\\
Then $\Rcat$ is a full reflective subcategory of $\Dlubpo$.\\
The adjunction is $\langle\Rcomp,\Rinc,\Reta\rangle \typ \Dlubpo \funad \Rcat$
with\\
$\Rcomp\typ \Dlubpo\fun \Rcat$ the functor mapping each 
$D\in \Dlubpo$ to $\Rcomp(D)$
and each $\fDE$ to the same $f\typ\Rcomp(D)\fun\Rcomp(E)$,\\
$\Rinc\typ \Rcat\fun\Dlubpo$ the inclusion functor,\\
$\Reta\typ \id_{\Dlubpo}\fun \Rinc\comp\Rcomp$ 
the natural transformation with\\
$\Reta_D\typ D\fun\Rinc(\Rcomp(D))$, $\Reta_D = \id$.
\end{thm}
\proof
$\Rcomp$ is a functor by proposition \ref{p:frcomp}.\\
$\Reta$ clearly is a natural transformation.\\
We have to prove that for every $D\in \Dlubpo$, $E\in\Rcat$,
$f\typ D\fun \Rinc(E)$,
there is a unique $g\typ\Rcomp(D)\fun E$ with $f=\Rinc(g)\comp \Reta_D$.
We choose for $g$ the same function $f$,
as $f\typ \Rcomp(D)\fun\Rcomp(E)=E$ is \cont\ by proposition \ref{p:frcomp}.
\qed

\begin{defi}\label{d:function}
Let $D,E$ be dlubpos.\\
We define the \defin{(pointwise) function space dlubpo} $\DexpoE$ by\\
$|\DexpoE |$ the set of all continuous $\fDE$,\\
$f\lex_{\DexpoE}g$ iff for all $x\in \Del$ it is $fx\lexE gx$,\\
if $F\sub\DexpoEel$ is directed and $f\in\DexpoEel$,
then $F\nlub_{\DexpoE} f$ iff for all $d\in \Del$ it is
$Fd = \set{gd}{g\in F} \nlubE fd$.

In this definition it is $\Lubex F=f$.
The singleton axiom is fulfilled.
\end{defi}

\begin{prop}\label{p:invariant}
Let $\Rcal$ be an invariant lub-rule class with singletons and $D,E \in \Rcat$. Then
\begin{enumerate}[(1)]
\item
$\terminal\in \Rcat$ is the categorical terminal object,
\item
$D\pro E\in \Rcat$ is the categorical product,
\item
$\DexpoE\in \Rcat$ (but need not be the categorical exponent).
\end{enumerate}
\end{prop}
\proof
\textbf{(1)}
On $(|\terminal|,\lex)$ the only valid lub-rules are
$S\ruleto \{\bot\}$ for $S\sub \{\emptyset,\{\bot\}\}$
and they are all fulfilled by $\terminal$.\\
\textbf{(2)}
Let $((|D\pro E|,\lex),P\ruleto A)$ be a derived directed lub-rule of $\Rcal$,
such that the elements of $P$ are natural.
(We have to show that $A$ is natural.)\\
Then the elements of $\fsset{\proi}P$ and $\fsset{\prot}P$ are
also natural in $D$ \resp $E$.\\
$((\Del,\lex),\fsset{\proi}P\ruleto \fset{\proi}A)$ and
$((\Eel,\lex),\fsset{\prot}P\ruleto \fset{\prot}A)$ are derived rules of $\Rcal$,
by invariance of $\Rcal$ \wrt the \cont\ functions $\proi$ \resp $\prot$.\\
Thus $\fset{\proi} A$ and $\fset{\prot}A$ are natural in $D$ \resp $E$.
Therefore $A$ is natural in $D\pro E$,
so $D\pro E$ fulfills the lub-rule $P\ruleto A$.
We get $D\pro E\in \Rcat$.\\
As $\Rcat$ is a full subcategory of $\Dlubpo$,
$\Rcat$ has the projections and pairing functions,
so $D\pro E$ is the categorical product in $\Rcat$.\\
\textbf{(3)}
Let $((|\DexpoE|,\lex),P\ruleto A)$ be a derived directed lub-rule of $\Rcal$,
such that the elements of $P$ are natural.
(We have to show that $A$ is natural.)\\
Then the elements of $Px=\set{Bx}{B\in P}$, with $Bx=\set{fx}{f\in B}$,
are also natural in $E$.\\
$((\Eel,\lex),Px\ruleto Ax)$ is a derived lub-rule of $\Rcal$,
by invariance of $\Rcal$ \wrt the function $\la f.fx$,
that respects the lubs of the $B\in P$.\\
Thus $Ax$ is natural in $E$ for every $x\in \Del$.
Therefore $A$ is natural in $\DexpoE$, 
so $\DexpoE$ fulfills the lub rule $P\ruleto A$.
We get $\DexpoE \in \Rcat$.
\qed

%% file: T6.tex
\section{The ccc of S10-dlubpos}

In this section we find a large cartesian closed full subcategory of $\Dlubpo$
whose exponents are the (pointwise) function spaces $\DexpoE$ which are known
from natural domains and from  definition \ref{d:function}.
The underlying structures are the S10-dlubpos,
which fulfill the singleton axiom and a new axiom S10.
All natural domains are also S10-dlubpos.

\begin{defi}
A dlubpo $D$ is an \defin{S10-dlubpo} if it fulfills the axiom:\\
\textbf{(S10)}
If $I$ is an index set with a directed partial order,\\
and $\yijII$ is a two-parametric family of elements of $\Del$ 
that is monotonic in each parameter $i$ and $j$,\\
and for all $i\in I$ it is $\yijIIj\nlubD x_i$ for some $x_i$,
and $\xiI\nlubD u$ for some $u$,\\
and for all $j\in I$ it is $\yijIIi\nlubD z_j$ for some $z_j$,
and $\zjI \nlubD u$,\\
then it is $\yiiI\nlubD u$.\\
$\Stdlubpo$ is the full subcategory of $\Dlubpo$ of all S10-dlubpos.
\end{defi}

The axiom looks like a cross-over of axioms S4(1\Right) and S4(2) of natural domains.
But note that all the natural convergences in  the axiom are directed,
by the monotonocity of the family.

\begin{prop}
The axiom S10 follows in a dlubpo from axioms (S4(1\Right) and S4(2)).
It also follows from axioms (S5 and S7(transitivity)).\\
So every natural domain
is an S10-dlubpo.
\qed
\end{prop}

\begin{prop}
The axioms of an S10-dlubpo (\ie S3(singleton) and S10)
form valid lub-rules and an invariant lub-rule class.\\
So every cdlubpo is an S10-dlubpo.
\end{prop}
\proof
This is clear for the singleton axiom S3.\\
An instance of axiom S10 translates to a lub-rule $(D,P\ruleto A)$
for a partial order $D$ with an $I$ and $\yijII$ in $D$ 
and where $P$ consists of $\yijIIj$ for all $i\in I$ (with lub $x_i$),
$\xiI$ (with lub $u$),
$\yijIIi$ for all $j\in I$ (with lub $z_j$),
$\zjI$ (with lub $u$).
$A$ is $\yiiI$.
$u$ is the lub of $A$.

Let $E$ be another partial order and $\fDE$ be monotonic and respecting the lubs of the 
elements of $P$.
In the same way it can be seen that $fu$ is the lub of $\fsetf A$,
so the lub-rule $(D,P\ruleto A)$ is valid.
It is also clear that $(E,\fssetf P\ruleto \fsetf A)$ is a lub-rule for an instance
of axiom S10,
so that the lub-rule class is invariant.
\qed

\begin{defi}
For every directed partial order $(I,\lex)$, 
that will be used as an index set,
we construct a dlubpo $\Ib =(|\Ib|,\lex_{\Ib},\nlub_{\Ib})$ 
as a completion of $I$ by a lub:\\
If $I$ contains an upper bound of itself,
then this will be called $t$ and $|\Ib|=I$ and $\lex_{\Ib}\; =\;\lex$.\\
Otherwise, $|\Ib|=I\join \{t\}$ with a new top element $t$,
and $i\lex_{\Ib} j$ iff
($i,j\in I$ and $i\lex j$)
or ($i\in |\Ib|$ and $j=t$).\\
In every case, the convergence is $\{x\}\nlub_{\Ib} x$ for $x\in |\Ib|$,
and $I\nlub_{\Ib} t$.
\end{defi}

\begin{lem}\label{l:evalS10}
Let $\Rcal$ be an invariant lub-rule class with singletons.\\
Let for every directed partial order $I$ be $\Is=\Rcomp(\Ib)$.\\
Let $E$ be a dlubpo generated by $\Rcal$.
Then the following are equivalent:
\begin{enumerate}[(1)]
\item
$E$ fulfills axiom S10.
\item
For every dlubpo $D$ it is $\eval\typ (\DexpoE)\pro D\fun E$ \cont.
\item
For every directed partial order $I$ it is $\eval\typ (\Is\expo E)\pro \Is\fun E$ \cont.
\item
For every dlubpo $D$ it is $(\DexpoE) = (\DexpoSE)$.
\end{enumerate}
\end{lem}
\proof
\textbf{(1)\Right (2)}
First, $\eval$ is monotonic.
Now let $A\sub (\DexpoE) \pro D$ be directed with $A\nlub (f,d)$.
We have to show that $\fset{\eval}A\nlub fd$.
It is $\fset{\eval}A=\set{(\proi x)(\prot x)}{x\in A}$.\\
Let $I=A$ be the index set for axiom S10.
For $i,j\in I$ let $\yij=(\proi i)(\prot j)$.\\
For $i\in I$ it is $\yijIIj\nlub(\proi i)d =: x_i$,
as $\{\prot j\}_{j\in I}\nlub d$ and as $\proi i$ is \cont.\\
It is $\xiI\nlub fd =: u$,
as $\{\proi i\}_{i\in I}\nlub f$ and therefore $\{\proi i\}_{i\in I}\, d\nlub fd$.\\
For $j\in I$ it is $\yijIIi\nlub f(\prot j) =: z_j$,
as $\{\proi i\}_{i\in I}\nlub f$.\\
It is $\zjI\nlub fd$, as $\{\prot j\}_{j\in I}\nlub d$.\\
Therefore the conditions of axiom S10 are fulfilled and $\yiiI\nlub fd$,
so $\fset{\eval} A\nlub fd$.

\textbf{(2)\Right (4)}
For $F\sub \DexpoEel$ directed and $f\in \DexpoEel$ it is
$F\nlub_{\DexpoE} f$ iff $F\nlub_{\DexpoSE} f$.

\textbf{(4)\Right (3)}
Follows directly from proposition \ref{p:evalstar}.

\textbf{(3)\Right (1)}
Let $I$ be an index set with a directed partial order,
and $\yijII$, $\xiI$, $\zjI$ and $u$ in $E$ as the axiom S10 describes.\\
For $i\in I$ we define functions $f_i\typ \Ib\fun E$ by
$f_ij=y_{ij}$ for $j\in I$,
and $f_it=x_i$.\\
$f_i$ is continuous, as $\yijIIj\nlub x_i$.\\
As $\Rcal$ is invariant,
$f_i$ is also a continuous function $f_i\typ \Is\fun E$, by proposition \ref{p:frcomp}.\\
We define $f_t\typ \Ib\fun E$ by $f_tj=z_j$ for $j\in I$,
and $f_tt=u$.\\
$f_t$ is continuous, as $\zjI\nlub u$.
Again, $f_t$ is also a continuous function $f_t\typ \Is\fun E$.\\
Let $A\sub |\Is\expo E|\pro|\Is|$ with $A=\set{(f_i,i)}{i\in I}$.
$A$ is directed.\\
We have $\{f_i\}_{i\in I}\nlub f_t$,
as (a) $\{f_ij\}_{i\in I}\nlub f_tj$ for $j\in I$, as $\yijIIi\nlub z_j$,\\
and as (b) $\{f_it\}_{i\in I}\nlub f_tt$, as $\xiI\nlub u$.\\
And we have $\{i\}_{i\in I}\nlub t$,
so $A\nlub (f_t,t)$.\\
As $\eval\typ (\Is\expo E)\pro \Is\fun E$ is \cont,
we get $\fset{\eval}A\nlub \eval(f_t,t)=u$,
so $\yiiI\nlub u$.
\qed

As a difference to the general function space $\DexpoSE$,
the curried functions for $\DexpoE$ are \cont:
\begin{prop}\label{p:currynormal}
Let $C,D,E$ be dlubpos.
For any $f\typ C\pro D\fun E$, $\curry(f)\typ C\fun(\DexpoE)$,
with $\curry(f)c = \la d\in D. f(c,d)$, is \cont\ (and has \cont\ results).
\end{prop}
\proof
Let $A\sub \Cel$ be directed with $A\nlub a$.\\
Then $\set{\la d\in D. f(x,d)}{x\in A}\nlub_{\DexpoE} \la d\in D. f(a,d)$,
as $f$ is \cont\ in its first argument.
\qed

\begin{thm}\label{t:S10}\hfill
\begin{enumerate}[(1)]
\item
Let $\Rcal$ be an invariant lub-rule class with singletons.\\
The subcategory $\Rcat$ of $\Dlubpo$ is a ccc with pointwise exponents $\DexpoE$\\
$\ifff$ every $E\in \Rcat$ fulfills axiom S10.\\
($\Rcat$ also has normal terminal and normal products.)
\item
From (1) follows that $\Stdlubpo$, and the category of natural domains,
and $\Cdlubpo$ are cccs with the pointwise exponent $\DexpoE$.
\item
$\Stdlubpo$ is the largest full sub-ccc of $\Dlubpo$ 
which is generated by an invariant lub-rule class and has the pointwise exponents $\DexpoE$.
\end{enumerate}
\end{thm}
\proof
(1) $\ifthen$:
By proposition \ref{p:expo1}(1) the evaluation function for the exponent in $\Rcat$ 
is the normal $\eval$.\\
For every directed partial order $I$ it is $\Is=\Rcomp(\Ib)$ in $\Rcat$.\\
Let $E\in \Rcat$.
It must be $\eval\typ (\Is\expo E)\pro \Is\fun E$ \cont,
so by lemma \ref{l:evalS10}(3\Right 1), $E$ fulfills S10.

(1) $\thenif$:
From proposition \ref{p:invariant} follows that $\terminal$ is in $\Rcat$ the terminal object,
$D\pro E$ is in $\Rcat$ the product, and $\DexpoE$ is in $\Rcat$.\\
That $\DexpoE$ is the exponent in $\Rcat$ follows from lemma \ref{l:evalS10}(1\Right 2)
and proposition \ref{p:currynormal}.

(2) and (3) follow immediately from (1).
\qed

%% file: T7.tex
\section{Example of a natural domain that is no cdlubpo}\label{s:incomplete}

We give the counter-examples of this section in two stages of increasing complexity.
The first example shows that the lub-rule class $\Scal$ corresponding
to the axioms of natural domains is not complete in the sense of Section \ref{s:lubrule}.
It is simpler, because it refers to general, non-directed natural subsets.

The second example shows a natural domain whose directed natural subsets do not form
a cdlubpo.
It is more complicated, because it has to refer to \emph{directed} natural subsets.
Of course, the second example alone would be enough for all.

\begin{thm}
The lub-rule class $\Scal$ corresponding to the axioms of natural domains
is not complete.
This means that there is a natural domain that does not fulfill axiom S9 (closure)
for general (not necessarily directed) subsets.
\end{thm}
\proof
Here is the example, a finite natural domain $D$:

\begin{pspicture}(-0.2,-0.7)(3.0,2.0)
\firsto{d}{\S}{d}
\slineon{d}{b}{\NE}{1}{\W}{b}
\slineo{b}{e}{\SE}{1}{\S}{e}
\slineo{e}{c}{\NE}{1}{\E}{c}
\slineon{b}{a}{\NE}{1}{\N}{a}
\sline{a}{c}
\end{pspicture}

It has as (non-directed) natural lubs $\{b,c\}\nlub a$,
$\{d,e\}\nlub b$ and all natural lubs that can be deduced
from these by the axioms of natural domains:
$\{b,c,d\}\nlub a$, $\{b,c,e\}\nlub a$, $\{b,c,d,e\}\nlub a$
and all trivial natural lubs where the lub is an element of the natural subset.

But by a complete axiom system the further natural lub $\{d,c\}\nlub a$ could be deduced,
because $a\in \cl_{\{\{b,c\},\{d,e\}\}} \{d,c\}$.
\qed

It might still be that the natural domains 
coincide with the cdlubpos,
the closed dlubpos with complete axiom system.
But we have the theorem:

\begin{thm}
There is a natural domain $D$ whose directed natural subsets do not form a cdlubpo.
\end{thm}
\proof
Here is the example,
an infinite natural domain $D$ that is no cdlubpo, see figure \ref{f:incomplete}.

\newcommand{\bcbc}[9]%
{\dlineo{#1}{\noden2}{\S}{1}{\W}{#2}
\slineon{\noden2}{\noden4}{\NEE}{1.5}{\E}{#4}%
\dlineo{\noden4}{\noden5}{\S}{1}{\E}{#5}%
\slineo{\noden5}{\noden6}{\S}{1}{\E}{#6}%
\nexton{\noden6}{\noden7}{\S}{1}{\E}{#7}%
\slineo{\noden7}{\noden3}{\SWW}{1.5}{\W}{#3}%
\sline{\noden2}{\noden3}%
\dlineo{\noden7}{\noden8}{\S}{1}{\E}{#8}%
\slineo{\noden8}{\noden9}{\S}{1}{\E}{#9}}

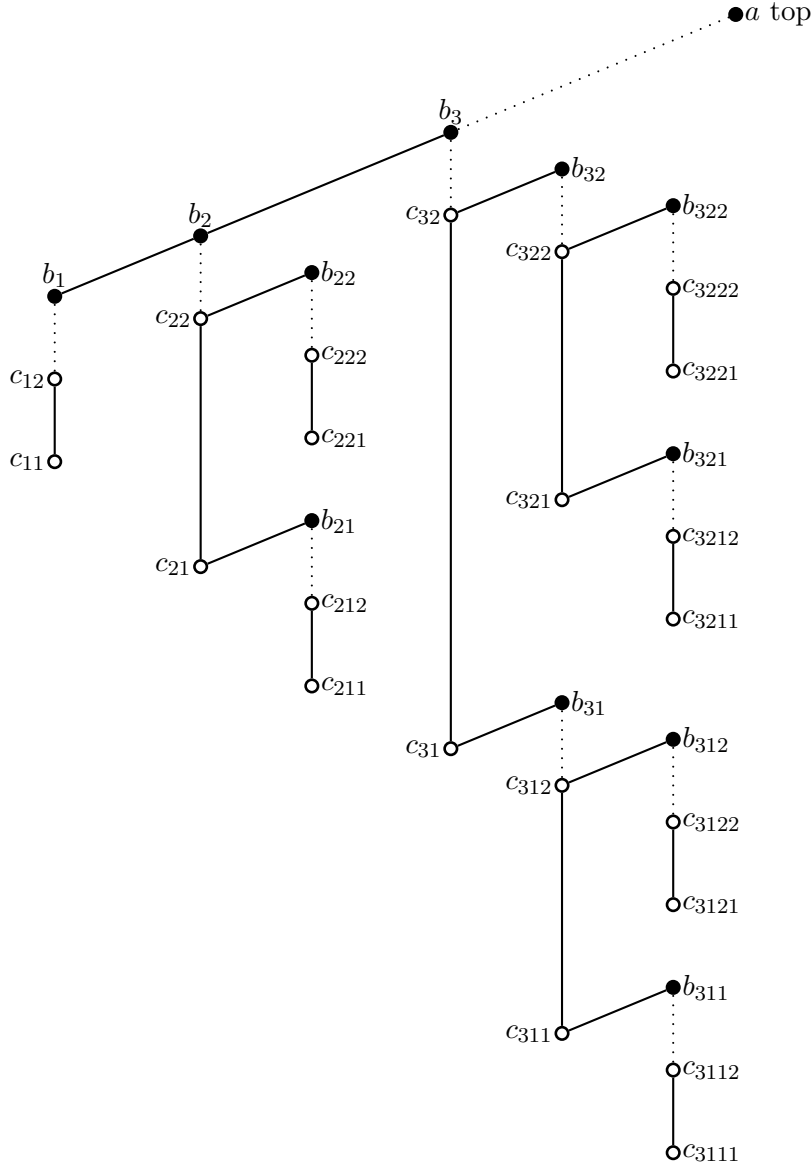
\begin{figure}[htb]
\begin{pspicture}(-0.7,-11.5)(10,4)
\firston{b1}{\N}{b_1}
\dlineo{b1}{c12}{\S}{1}{\W}{c_{12}}
\slineo{c12}{c11}{\S}{1}{\W}{c_{11}}
\slineon{b1}{b2}{\NEE}{2}{\N}{b_2}
\newcommand{\noden}{d}
\bcbc{b2}{c_{22}}{c_{21}}{b_{22}}{c_{222}}{c_{221}}{b_{21}}{c_{212}}{c_{211}}
\slineon{b2}{b3}{\NEE}{3.5}{\N}{b_3}
\dlineo{b3}{c32}{\S}{1}{\W}{c_{32}}
\slineon{c32}{b32}{\NEE}{1.5}{\E}{b_{32}}
\renewcommand{\noden}{e}
\bcbc{b32}{c_{322}}{c_{321}}{b_{322}}{c_{3222}}{c_{3221}}{b_{321}}{c_{3212}}{c_{3211}}
\slineo{c32}{c31}{\S}{7}{\W}{c_{31}}
\slineon{c31}{b31}{\NEE}{1.5}{\E}{b_{31}}
\renewcommand{\noden}{f}
\bcbc{b31}{c_{312}}{c_{311}}{b_{312}}{c_{3122}}{c_{3121}}{b_{311}}{c_{3112}}{c_{3111}}
\dlineon{b3}{a}{\NEE}{4}{\E}{a \text{ top}}
\end{pspicture}
\caption{natural domain $D$ (partial)}\label{f:incomplete}
\end{figure}

The elements of $D$ are:\\
$a$ the top element (\ie $x\lex a$ for all $x\in \Del$).\\
The elements of $B=\{b_i\}_{i\geq 1}$ with $b_i\lex b_j$ for $i\leq j$, it is $B\nlub a$.\\
For every $n\geq 1$ there are the elements $b_{nw}$,
where $w$ is a word of numbers $\geq 1$ with $1\leq \length(w)<n$,\\
and the elements $\cnw$, where $w$ is a word of numbers $\geq 1$ with $1\leq \length(w)\leq n$.\\
(Words of numbers will always be denoted by $w$.)\\
It is $c_{w1}\lex c_{w2}\lex \dots \lex b_w$
and $\{c_{w1}, c_{w2},\dots\}\nlub b_w$ and $c_w\lex b_w$ for all valid words $w$.\\
Let $C=\set{\cnw}{n\geq 1, \length(w)=n}$,
these are the $\cnw$ in the rightmost position under each $b_n$.
$C$ is not directed, it has lub $a$.\\
Like the example in section \ref{s:sazonov}, we erect an ``artificial'' directed set $\Cb$ on $C$,
which is not depicted in the diagram:\\
Let $\Cb$ be the set of all finite subsets of $C$.
All elements of $\Cb$ are also elements of $D$.\\
For all $A,B\in \Cb$: $A\lexD B$ iff $A\sub B$, and of course $A\lexD a$.\\
For all $x\in C$: $x\lexD \{x\}\in \Cb$.
$\Cb$ is directed with lub $a$.\\
The natural domain $D$ is the reflexive, transitive closure of the order $\lex$ and the
closure of $\nlub$ under the axioms of natural domain.

The intuition of the example is this:
To deduce $C\nlub a$ (and $\Cb\nlub a$) one would start with $B\nlub a$ and stepwise
replace in $B$ elements $b_w$ by $\{c_{wi}\}_{i\geq 1}$
and elements $c_w$ by $b_w$,
working from left to right under each of the $b_i$,
until we reach the elements of $C$ and have replaced all other elements.
Under $b_i$ this takes at least $2i-1$ sequential steps,
so no deduction can do this for all $b_i$.
(There is no separate deduction of natural lubs $b_i$.
$X\nlub b_i$ for $X\sub \{\cij\}_{j\geq 1}$ infinite are the only non-trivial natural lubs $b_i$.)
Therefore we have:

\begin{prop}
In $D$ it is not $C\nlub a$ and not $\Cb\nlub a$.
\end{prop}
\proof
In the following ``non-trivial natural lub'' means that the natural subset does not contain its own lub.
The only non-trivial natural lubs $\neq a$ are the $C'\nlub b_w$ with a valid word $w$ and 
where $C'$ contains an infinite subset of $\{c_{wj}\}_{j\geq 1}$.
This is so because in  such a natural set no $c_{wj}$
can be replaced by $b_{wj}$, as it is not $b_{wj}\lex b_w$.

\medskip
We will prove that every non-trivial $A\nlub a$ fulfills this \emph{condition cond}:\\
There is some $n\geq 0$, and an infinite set $N$ of numbers $\geq 1$,
and a map $d\typ N\fun A$,
such that for every $i\in N$ it is $d(i)=\biw$ or $d(i)=\ciw$ for some word $w$ with $\length(w)\leq n$.
\medskip

Here is the proof:\\
Every $A\nlub a$ must be deduced in  a chain 
$B=A_1\under A_2\under \dots \under A_m = A$,
where $A_i\nlub a$ is non-trivial.
See the definition \ref{d:under} of $\under$ and proposition \ref{p:under}
and note that $a$ is the top element.\\
$B$ fulfills the condition cond.
We have to prove that if $A$ fulfills the condition cond, then also $A'$ with $A\under A'$.

A step $A\under A'$ means the following:
\begin{enumerate}[(1)]
\item
For every $b_i\in A$ (there is some $b_j\in A'$ with $j\geq i$)
or (there is $C'\sub A'$ with an infinite $C'\sub \{\cij\}_{j\geq 1}$).\\
\item
For every $\biw\in A$, with $\length(w)\geq 1$, (there is $\biw\in A'$)
or (there is $C'\sub A'$ with an infinite $C'\sub \{\ciwj\}_{j\geq 1}$).\\
\item
For every $\cij\in A$, (there is some $\cik\in A'$ with $k\geq j$)
or (there is some $b_k\in A'$ with $k\geq i$)
or $\bij \in A'$).\\
\item
For every $\ciwj\in A$, with $1\leq \length(w)\leq i-2$,
(there is some $\ciwk\in A'$ with $k\geq j$)
or ($\biw\in A'$) or ($\biwj\in A'$).\\
\item
For every $\ciwj\in A$, with $\length(w)=i-1$,
(there is some $\ciwk\in A'$ with $k\geq j$)
or ($\biw\in A'$)
or (there is some $F\in A'$ with $F\in \Cb$ and $\ciwj\in F$).\\
\item
For every $F\in A$, with $F\in \Cb$,
there is $F'\in A$ with $F'\in \Cb$ and $F\lex F'$.
\end{enumerate}

We prove that $A'$ fulfills the condition cond. There are two cases:\\
(a) $A'\cap B$ is infinite:\\
Then $A'$ fulfills the condition cond with $n'=0$,
$N'=\set{i}{b_i\in A'\cap B}$, $d'(i)=b_i$.\\
(b) otherwise:\\
Then $A'$ fulfills the condition cond with the following data:\\
$n'=n+1$.\\
$N'$ is the set of all $i\in N$ with $i\geq n+1$ that fulfill the following two conditions:
\begin{itemize}
\item If $d(i)=b_i$, then $b_i\in A'$ or some $\cij\in A'$.
\item If $d(i)=\cij$, then (some $\cik\in A'$ with $k\geq j$)
or ($\bij\in A'$) or ($b_i\in A'$).
\end{itemize}

(These conditions rule out the cases of $i$ where $b_i$ or $\cij$ is replaced in $A'$
only by a $b_k$ with $k>i$.
$N'$ is still infinite as the $i\in N$ with $i\geq n+1$ are infinitely many
and the ruled out cases of $i$ can only be finitely many,
otherwise we would have case (a).)

$d'(i)$ is defined for $i\in N'$ as follows:
\begin{itemize}
\item
If $d(i)=\biw$:
if $\biw\in A'$ then $d'(i)=\biw$ else $d'(i)=\ciwj$ for some $j$ such that $\ciwj\in A'$.
\item
If $d(i)=\ciwj$:
if there is some $\ciwk\in A'$ then $d'(i)=\ciwk$\\
else (if $\biw\in A'$ then $d'(i)=\biw$ else $d'(i)=\biwj$).\\
(Here it is $i\geq n+1$ and $\length(wj)\leq n$, so $\length(w)<i-1$.
Therefore $\ciwj$ is not replaced in $A'$ by some $F\in \Cb$.)
\end{itemize}

It is clear from the meaning of the step $A\under A'$ above and the choice of $N'$ 
that $d'(i)$ is an element of $A'$.
The word length of the index of $d'(i)$ stays the same as for $d(i)$ or increases by $1$.

So we have proved that every non-trivial $A\nlub a$ fulfills the condition cond.
This proves that $C\nlub a$ and $\Cb\nlub a$ are not fulfilled in $D$.
\qed

\begin{prop}
From the specified natural lubs of $D$ it can be deduced by the closure axiom S9
that $\Cb\nlub a$.
\end{prop}
\proof
Let $\Cs=\clP(\Cb)$, where $P$ is the specified set of natural sets of $D$.
We have to show that $a\in \Cs$.\\
First, for every $c\in C$ it is $c\in\Cs$, as $c\lexD \{c\}\in \Cb$.\\
For all $n\geq 1$, we now work under $b_n$:\\
We have that all $c_{nw}\in \Cs$ with $\length(w)=n$, these are the elements from $C$.\\
If for some $1\leq k\leq n$, all $\cnw\in \Cs$ with $\length(w)=k$,
then all $b_{nw}\in \Cs$ with $\length(w)=k-1$.\\
If for some $1\leq k\leq n-1$, all $b_{nw}\in \Cs$ with $\length(w)=k$,
then all $\cnw\in\Cs$ with $\length(w)=k$.\\
As a result of this consecutive process,
we get $b_n\in \Cs$ for all $n\geq 1$, so $a\in \Cs$.
\qed
From the two propositions it is immediate that $D$ is not a cdlubpo.
\qed

\begin{rem}
Note that $D$ of the theorem is not a (naturally) algebraic natural domain
in the sense of definition 3.2 of \Saz.
In the next section we show that every algebraic natural domain is a cdlubpo.
\end{rem}
\begin{rem}
Another question is that for an augmentation of the axioms of natural domains
so that they describe cdlubpos.
This could be a modification of axiom S8 (under)
so that the ``limit'' $L$ of an ascending $\under$-chain $\lex a$
leads to a natural lub $L\nlub a$.
But this would need a complicated definition of ``limit'',
so we prefer our axiom S9 (closure) instead.
\end{rem}

%% file: T8.tex
\section{Algebraic dlubpos}

We adapt the definitions of (naturally) finite elements and (naturally) algebraic natural domains
from Sazonov \Saz\ to the case of dlubpos.
We show that an algebraic dlubpo that fulfills axiom S6 (cofinality) is a cdlubpo.
This proves that (naturally) algebraic natural domains and algebraic cdlubpos are the same.

\begin{defi}[Sazonov, definitions 3.1, 3.2 of \Saz]\label{d:alg}
Let $D$ be a dlubpo.\\
An element $a\in\Del$ is \defin{finite} if for every (directed) $B\nlub b$ with
$a\lex b$ there is $b'\in B$ with $a\lex b'$.
$\Do$ is the set of finite elements of $D$.\\
For $d\in\Del$ we define $\downo d=\set{a\in \Do}{a\lex d}$.\\
$D$ is \defin{algebraic} if for every $d\in\Del$ it is $\downo d\nlub d$.
(This includes that $\downo d$ is directed.)
\end{defi}
Note that we omit Sazonov's prefix ``naturally'' (finite, algebraic)
as this is given by the fact that we are talking about dlubpos.

The next lemma shows that the finiteness property extends to closures.

\begin{lem}\label{l:fincl}
Let $D$ be a dlubpo.\\
Let $A\sub\Del$, $a\in \Do$ and $a\in \clD A$.
Then there is some $a'\in A$ with $a\lex a'$.
\end{lem}
\proof
Let $(N,r,\lab,\pre)$ be a deduction of $a$ from $A$ in the cl-rule system.
We construct a path 
$r=n_1, n_2\in \pre n_1, n_3\in \pre n_2,\dots, n_k$
from the root $r$ to a leaf $n_k$ inductively as follows.
It is always $a\lex \lab n_i$.\\
If $\lab n_i=b$ and $\pre n_i=\{m\}$ with $b\lex \lab m$,
then we choose $n_{i+1}=m$.\\
If $\lab n_i=b$ and $\pre n_i=M$ with $\fset{\lab} M\nlub b$,
then there is some $m\in M$ with $a\lex \lab m$, as $a$ is finite.
We choose $n_{i+1}=m$.\\
By well-foundedness of the deduction, this process ends with a leaf $n_k$ 
with $\lab n_k=a'\in A$ and $a\lex a'$.
\qed

There is an interesting property of dlubpos connected to algebraicity:

\begin{defi}
Let $D$ be a dlubpo.
$D$ is \defin{finite-determined}
if there is a subset $F\sub \Del$ such that for every directed $A\sub\Del$ with lub $a$ it is:\\
$A\nlub a  \ifff$ for all $b\in F$ with $b\lex a$ there is some $a'\in A$ with $b\lex a'$.
\end{defi}

The elements of $F$ are finite in the sense of definition \ref{d:alg}.
Every algebraic dlubpo $D$ fulfills the direction $\ifthen$ with $F=\Do$.
But a finite-determined $D$ need not be algebraic,
as it is not stipulated that every element $a$ of $D$ is a (natural) lub of the directed set
of finite elements below $a$.
On the other side our definition demands more,
namely the direction $\thenif$,
which is not entailed by algebraic dlubpos.

\begin{prop}
If $D$ is a finite-determined dlubpo and for every $a\in \Del$ it is 
$a=\Lubex \downo a$ directed,
then $D$ is algebraic.
\qed
\end{prop}

Remember that the axiom S6 (cofinality) for directed subsets is:\\
If $X,Y\sub\Del$ directed, $X\nlub x$ and $X\lset Y\lset x$, then $Y\nlub x$.

\begin{prop}[Reinhold Heckmann, personal communication]\label{p:algfin}
\hfill\\
If $D$ is an algebraic dlubpo with axiom S6 for directed subsets,
then $D$ is finite-determined with $F=\Do$.
\end{prop}
\proof
We must prove the direction $\thenif$ of the definition.\\
The right side means: $\downo a\lset A$.
It is $A\lset a$ and $\downo a\nlub a$.\\
By axiom S6 we get $A\nlub a$.
\qed

\begin{prop}\label{p:fincd}
If $D$ is a finite-determined dlubpo,
then $D$ is a cdlubpo.
\end{prop}
\proof
Let $D$ be finite-determined by the subset $F$.\\
Let $A\sub \Del$ be directed with lub $a$ and $a\in \clD A$.
We have to show $A\nlub a$.\\
Let $b\in F$ with $b\lex a$.
Then $b\in \Do$ and $b\in \clD A$.\\
By lemma \ref{l:fincl} there is some $a'\in A$ with $b\lex a'$.
\qed

\begin{thm}\hfill
\begin{enumerate}[(1)]
\item
If $D$ is an algebraic dlubpo with axiom S6 (cofinality) for directed subsets,
then $D$ is a cdlubpo.
\item
For any dlubpo $D$ the following are equivalent:\\
(a) $D$ is an algebraic dlubpo that fulfills axiom S6 for directed subsets.\\
(b) $D$ is an algebraic cdlubpo.\\
(c) $D$ is an algebraic natural domain.
\end{enumerate}
\end{thm}
\proof
(1) follows from propositions \ref{p:algfin} and \ref{p:fincd}.\\
(2) is clear.
\qed

%% file: T9.tex
\section{Restricted partial orders and restricted dcpos}

We want to establish cdlubpos as the most general canonical structures in which to build
non-complete programming language models,
corresponding to dcpos as the most general structures to build complete models.
(These structures still lack algebraicity/continuity.)
We have already seen that cdlubpos are just the dlubpos that are generated by a complete
lub-rule class.
In this section we see another characterization of cdlubpos that shows their canonicity:
they are just the dlubpos that are realized by ``restricted partial orders'' (rpos),
in which they are embedded.
This embedding can also be in a ``restricted dcpo'' (rdcpo),
which is a dcpo with a subset designated as the set of ``proper'' elements.

But before describing the rdcpos we give a property that is sufficient for
dlubpos to be cdlubpos.

\begin{prop}\label{p:detK}
Let $D$ be a dlubpo and $K$ be a subclass of the class of all 
dlubpo morphisms with domain $D$.
(The morphisms of $K$ could be given by a subcategory in which $D$ lives.)\\
We say that \defin{``the natural lubs of $D$ are determined by $K$''}\\
if for all directed $A\sub \Del$ with lub $a$ it is:\\
if all $\fDE$ in $K$ respect the lub of $A$ (\ie $fa=\Lubex \fsetf A$),
then $A\nlubD a$.\\
In this case $D$ is a cdlubpo.
\end{prop}
\proof
Let $((\Del,\lexD),P\ruleto A)$ be a valid directed lub-rule
and let all elements of $P$ be natural in $D$.
All $\fDE$ in $K$ respect the lubs of the elements of $P$,
so also respect the lub of $A$, as the lub-rule is valid.\\
Therefore $A$ is natural in $D$ by the definition,
so $D$ fulfills the lub-rule.
Thus $D$ is a cdlubpo.
\qed

We now come to the realization of cdlubpos by restricted partial orders and restricted dcpos.

\begin{defi}
A structure $E=(\Eel,\real{E},\lexE)$ is a \defin{restricted partial order (rpo)} if\\
$\Eel$ is a set of (proper) \defin{elements},\\
$\real{E}$ is a set of \defin{realizers}, with $\Eel \sub \real{E}$,\\
$\lexE$ is a partial order on $\real{E}$.\\
We define $\breal{E}=\real{E}\setminus\Eel$, the set of \defin{blind realizers}.
\end{defi}

The idea of a partial order realized by a cpo appears in \cite{Simpson},
but there every realizer realizes exactly one proper element (there are no blind realizers),
and a proper element may be realized by several realizers.

\begin{defi}
An rpo $E$ \defin{realizes} a dlubpo $D$ \defin{by} an order isomorphism\\
$\vphi\typ (\Del,\lexD)\fun (\Eel,\lexE)$
if for all directed $\Lubex A=a$ in $D$:
\[ A\nlubD a \ifff \vphi a = \Lubex \fset{\vphi} A. \]
\end{defi}

\begin{prop}\label{p:rpo}\hfill
\begin{enumerate}[(1)]
\item
If an rpo $E$ realizes a dlubpo $D$, then $D$ is a cdlubpo.
\item
Every cdlubpo $D$ is realized by the rpo $(\fset{\inn}\Del,\Dhat,\lex)$ by $\inn\typ\Del\fun\fset{\inn}\Del$,
see def.\ \ref{d:cl}.
\item
The cdlubpos are exactly the dlubpos realized by rpos.
\end{enumerate}
\end{prop}
\proof
\textbf{(1)}
This is essentially the proof of proposition \ref{p:detK},
when we construe $E$ as a dlubpo with all directed lubs natural.\\
Let $\vphi$ be the embedding of $D$ in $E$.
Let $((\Del,\lex),P\ruleto A)$ be a valid directed lub-rule such that the elements 
of $P$ are natural in $D$.
$\vphi$ respects the lubs of the elements of $P$.
As the lub-rule is valid,
$\vphi$ respects also the lub of $A$,
so $\vphi(\Lubex A)=\Lubex\fset{\vphi} A$.\\
As $E$ realizes $D$ by $\vphi$, it is $A\nlub a$ in $D$.
So $D$ fulfills all valid directed lub-rules, it is a cdlubpo.\\
\textbf{(2)}
Let $A\nlub a$ in $D$.
Then $\innD a = \Lubex(\fset{\innD}A)=\clD A$ by proposition \ref{p:in}.\\
For the reverse: 
It is $a\in\clD A$, therefore $A\nlub a$.
\qed

We want to realize our dlubpos in dcpos which can be constructed by a completion process.
We define the category of restricted dcpos:

\begin{defi}
A structure $E=(\Eel,\real{E},\lexE)$ is a \defin{restricted dcpo (rdcpo)}
if it is an rpo and $(\real{E},\lexE)$ is a dcpo.\\
Let $D,E$ be rdcpos.
A function $f\typ\Dreal\fun\Ereal$ is \defin{continuous}
if it is \cont\ on the dcpos $(\Dreal,\lexD)$ and $(\Ereal,\lexE)$,
and $\fsetf\Del\sub\Eel$.
In this case we write $\fDE$.\\
$\Rdcpo$ is the category of all rdcpos and \cont\ functions,
with normal function composition and identity functions.
\end{defi}

\begin{prop}
The cdlubpos are exactly the dlubpos realized by rdcpos.
\end{prop}
\proof
Follows from proposition \ref{p:rpo}, as $\Dhat$ is a dcpo.
\qed

There is an adjunction between
the categories $\Dlubpo$ and $\Rdcpo$,
which establishes an adjoint equivalence between $\Cdlubpo$ and a full
subcategory $\Crdcpo$ (closed rdcpos) of $\Rdcpo$.
The details and further developments in this direction will be found in a sequel paper.

\section{Outlook}\label{s:outlook}

We have introduced lub-rule classes and closed dlubpos corresponding to
complete lub-rule classes as a canonical alternative to natural domains.
Closed dlubpos can also be characterized as the dlubpos realized by restricted dcpos.
We will explore this connection with rdcpos further in a sequel paper.
It will turn out that there is an adjunction between the categories
which permits the transfer of much of the theory of cccs of algebraic dcpos
to closed rdcpos \resp closed dlubpos.

The different approaches to realize partial orders by dcpos,
Simpson's \cite{Simpson} and ours,
call for a common generalization:
in the new concept a proper element may be realized by \emph{several} realizers (Simpson),
and a realizer may realize one or none element (as here).
This would be a category of dcpos that carry also some kind of \emph{partial} preorder,
the correct definition of it is not yet clear.
(Simpson has a preorder.)

We indicate further directions of research.
The question of the topology of non-complete domains is not yet fully answered,
Sazonov made some first observations on this in \Saz.

In the introduction we posed as an open problem to find categories of abstract
incomplete domains with the existence of fixpoints of endofunctions.
Also interesting is the question to find such categories of concrete incomplete domains,
\ie domains based on mechanisms like games.